\documentclass[journal, twoside]{IEEEtran}

\usepackage{graphicx}
\usepackage{color}
\usepackage{cite}

\usepackage{acronym}
\usepackage{amsfonts}  
\usepackage{amsmath}
\usepackage{amssymb}
\usepackage{bm}   
\usepackage{subfigure}
\usepackage{multicol,multirow}
\usepackage{graphics} 
\usepackage{algorithm,algorithmic}

\newcommand{\bfx}{{\mbox{\boldmath $x$}}}

\newcommand{\Ts}{{\mbox{$\mathrm{T}_s$}}}
\newcommand{\btheta}{{\mbox{$\bm{\theta}$}}}

\acrodef{ADC}{Analog-to-Digital Converter}
\acrodef{AMPA}{$\alpha$-amino-3-hydroxy-5-methyl-4-isoxazolepropionic}
\acrodef{AR}{autoregressive}
\acrodef{CKF}{cubature Kalman filter}
\acrodef{CNS}{Central Nervous System}
\acrodef{DSS}{discrete state-space}
\acrodef{EKF}{extended Kalman filter}
\acrodef{GABA}{$\gamma$-aminobutyric acid}
\acrodef{HMM}{Hidden Markov Model}
\acrodef{QKF}{quadrature Kalman filter}
\acrodef{KF}{Kalman filter}
\acrodef{ODE}{ordinary differential equation}
\acrodef{OU}{Ornstein-Uhlenbeck}
\acrodef{MCMC}{Markov-Chain Monte-Carlo}
\acrodef{PCRB}{Posterior Cram\'er-Rao Bound}
\acrodef{PF}{particle filtering}
\acrodef{PMCMC}{Particle Markov-Chain Monte-Carlo}
\acrodef{RAM}{Robust Adaptive Metropolis}
\acrodef{RMSE}{root mean square error}
\acrodef{SIS}{Sequential Importance Sampling}
\acrodef{SMC}{Sequential Monte-Carlo}
\acrodef{SPKF}{sigma--point Kalman filter}
\acrodef{SRUKF}{square--root unscented Kalman filter}
\acrodef{SRCKF}{square--root cubature Kalman filter}
\acrodef{SRQKF}{square--root quadrature Kalman filter}
\acrodef{SRKF}{square--root Kalman filter}
\acrodef{SRSPKF}{square--root sigma--point Kalman filter}
\acrodef{SS}{state-space} \acrodef{UKF}{unscented Kalman filter}
\acrodef{UT}{unscented transform}

\begin{document}

\title{Sequential estimation of intrinsic activity and synaptic input in single neurons by particle filtering with optimal importance density}

\author{Pau Closas,~\IEEEmembership{Senior Member,~IEEE,} and Antoni
Guillamon
    \thanks{Copyright (c) 2015 IEEE. Personal use of this material is permitted. However, permission to use this material for any other purposes must be obtained from the IEEE by sending a request to pubs-permissions@ieee.org.}
    \thanks{P. Closas is with the Centre Tecnol\`{o}gic de Telecomunicacions de Catalunya (CTTC),
            Parc Mediterrani de la Tecnologia, Av. Carl Friedrich Gauss 7,
            08860 Castelldefels, Barcelona (Spain). (e-mail: pau.closas@cttc.cat)}
    \thanks{A. Guillamon is with the Department of Mathematics at the Universitat Polit\`{e}cnica de Catalunya (UPC),
            Av. Dr. Mara\~non, 44--50, 08028 Barcelona (Spain). (e-mail: antoni.guillamon@upc.edu)}
    \thanks{P. Closas is supported by the Generalitat de Catalunya under grant 2014-SGR-1567. A. Guillamon is supported by the MINECO grant MTM2012-31714 (DACOBIANO) and the Generalitat de Catalunya grant AGAUR 2014-SGR-504.}
    \thanks{This paper has supplementary downloadable material available at http://ieeexplore.ieee.org , provided by the author. The material includes a dynamic visualization of the estimation method.}
    }

\markboth{submitted to IEEE Trans. on Sel. Topics in Signal
Processing}{Closas and Guillamon: Sequential estimation of intrinsic
activity and synaptic input by particle filtering}

\maketitle

\begin{abstract}
This paper deals with the problem of inferring the signals and
parameters that cause neural activity to occur. The ultimate
challenge being to unveil brain's connectivity, here we focus on a
microscopic vision of the problem, where single neurons (potentially
connected to a network of peers) are at the core of our study. The
sole observation available are noisy, sampled voltage traces
obtained from intracellular recordings. We design algorithms and
inference methods using the tools provided by stochastic filtering,
that allow a probabilistic interpretation and treatment of the
problem. Using particle filtering we are able to reconstruct traces
of voltages and estimate the time course of auxiliary variables. By
extending the algorithm, through PMCMC methodology, we are able to
estimate hidden physiological parameters as well, like intrinsic
conductances or reversal potentials. Last, but not least, the method
is applied to estimate synaptic conductances arriving at a target
cell, thus reconstructing the synaptic excitatory/inhibitory input
traces. Notably, these estimations have a bound-achieving
performance even in spiking regimes.
\end{abstract}


\IEEEpeerreviewmaketitle

\section{Introduction}
\label{intro}

\PARstart{N}{euroscience} is the science that delves into the
understanding of the nervous system. It is one of the most
interdisciplinary sciences, gathering together experts from a vast
variety of fields of knowledge. Neuroscience is a rather broad
discipline and encompasses many aspects related to the \ac{CNS}. The
different topics in neuroscience can be studied from various
perspectives depending on the prism used to focus the problem. This
ranges from understanding the internal mechanisms that cause spiking
of a single cell (a neuron), to explaining the dynamics
occurring in populations of neurons that are interconnected. Even
more macroscopically, one could consider the analysis of functional
parts in the brain which are ultimately composed of individual
neurons grouped together. In this work, we are interested in the
microscopic view of the problem.

Measurements of membrane potential traces constitute the main
observable quantities to derive a biophysical neuron model. In
particular, the dynamics of auxiliary variables and the model
parameters are inferred from voltage traces, in a costly process
that typically entails a variety of channel blocks and clamping
techniques (see for instance \cite{Brette12}), as well as some
uncertainty in the parameter values due to noise in the signal.
Recent works in the literature deal with the problem of inferring
hidden parameters of the model, see for instance
\cite{Huys09,DitlevsenSamson2014,Lankarany2014} and, for an exhaustive
review, \cite{DitlevsenSamson2015}. In the same line, it is worth to
mention attempts to extract connectivity in networks of neurons from
calcium imaging \cite{Mishchenko2011}.

Apart from inferring intrinsic parameters of the model, voltage
traces are also useful to obtain valuable information about synaptic
input, an inverse problem with some satisfactory (see for instance
\cite{Rudolph2004,Pospischil08,Bedard11,Kobayashi11,Paninski12,Berg2013,Lankarany2013a})
but no complete solutions yet. The main shortcomings are the
requirement of multiple (supposedly identical) trials for some
methods to be applied, and the need of avoiding signals obtained
when ionic currents are active. The latter constraint arises from
the fact that many methods rely on the linearity of the signal and
this is not possible to achieve under quite general situations, like
spiking regimes (see \cite{Guillamon2006}) or subthreshold regimes
when specific currents (e.g., AHP, LTS, etc.) are active (see
\cite{Vich2015}).

An ideal method should be able to sequentially infer the time-course
of the membrane potential and its intrinsic/extrinsic activity from
noisy observations of a voltage trace. The main features of the
envisaged algorithm are fivefold $i)$ \emph{Single-trial:} the
method should be able to estimate the desired signals and parameters
from a single voltage trace, thus avoiding the experimental
variability among trials; $ii)$ \emph{Sequential:} the algorithm
should provide estimates each time a new observation is recorded,
thus avoiding re-processing of all data stream each time; $iii)$
\emph{Spike regime:} contrary to most solutions operating only under
the subthreshold assumption, the method should be able to operate in
the presence of spikes as well; $iv)$ \emph{Robust:} if the method
is model-dependent, thus implying knowledge of the model parameters,
then the algorithm should be provided with enhancements to
adaptively learn these parameters; and $v)$ \emph{Efficient:} the
performance of the method should be close to the theoretical lower
bounds, meaning that the estimation error cannot be substantially
reduced. Notice that the focus here is not on reducing the
computational cost of the inference method and thus we allow
ourselves to use resource-consuming algorithms. Indeed, the target
application does not demand (at least as a main requirement)
real-time operation, and thus we prioritized performance (i.e.,
estimation accuracy and the rest of features described earlier) in
our developments.

According to the above desired features, in this work, that
substantially extends our previous contributions
\cite{Closas13,Closas13b}, we are interested in methods that can
provide on-line estimation and avoid the need of repetitions that
could be contaminated by neuronal variability. Particularly, we
concentrate on methods to extract intrinsic activity of ionic
channels, namely the probabilities of opening and closing ionic
channels, and the contribution of synaptic conductances. We built a
method based on Bayesian theory to sequentially infer these
quantities from single-trace, noisy membrane potentials. The
material therein includes a discussion of the discrete state-space
representation of the problem and the model inaccuracies due to
mismodeling effects. We present two sequential inference algorithms:
$i$) a method based on \ac{PF} to estimate the time-evolving states
of a neuron under the assumption of perfect model knowledge; and
$ii$) an enhanced version where model parameters are jointly
estimated, and thus the rather strong assumption of perfect model
knowledge is relaxed. We provide exhaustive computer simulation
results to validate the algorithms and observe that they are
attaining the theoretical lower bounds of accuracy, which are
derived in the paper as well.

The results show the validity of the approach and its statistical
efficiency. Although we used for convenience a specific neuron model
(i.e., the Morris-Lecar) in the computer simulations, the proposed
procedure can be applied to any neuron model without loss of
generality.

In this paper we use the powerful tools of \ac{PF} to make
inferences in general state-space models. \ac{PF} are a set of
methods able to sample from the marginal filtering distribution in
situations where analytical solutions are hard to work out, or
simply impossible. In the recent years \acp{PF} played an important
role in many research areas such as signal detection and
demodulation, target tracking, positioning, Bayesian inference,
audio processing, financial modeling, computer vision, robotics,
control or biology
\cite{SeqMC01,SpecialIssue02,DjuricM03,Arulampalam02,Chen03,BeyondKF,Sarkka13book}.
At a glance, \ac{PF} approximate the filtering distribution of
states given measurements by a set of random points, properly
weighted according to the Bayes' rule. The generation of the random
particles can be done through a variety of distributions, known as
importance density. Particularly, we formulate the problem at hand
and observe that it is possible to use the optimal importance
density \cite{Doucet00}. This distribution generates particles close
to the target distribution and thus it can be shown to reduce the
variance of the particles. As a consequence, for a fix number of
particles, usage of this approach (not always possible) leads to
better accuracy results than other choices. To the author's
knowledge, the utilization of such sampling distribution is novel in
the context of neural model filtering. Similar works have used
\ac{PF} to track neural dynamics, but with no optimal importance
density (see \cite{Huys09,DitlevsenSamson2014}) or to estimate synaptic input from
subthreshold recordings \cite{Paninski12}, as opposite to our
proposed approach where we aim at providing estimates during the,
highly nonlinear, spike regime. These references use the expectation-maximization algorithm to estimate the model parameters. Lighter filtering methods based on
the Gaussian assumption were considered in the literature (see
\cite{Ullah09,Lankarany2013a,Berg2013} for instance), but the
assumption might not hold in general. For instance, due to outliers
in the membrane measurements or if more sophisticated models for the
synaptic conductances are considered. In these situations, a \ac{PF}
approach seems more appropriate. As mentioned earlier, the focus
here is on highly efficient and reliable filtering methods, rather
than on computationally light inference methods.

The remainder of the article is organized as follows. Section
\ref{sect:physiology} provides a brief overview of the main
biological concepts regarding neural activity, which are relevant to
this work. This section has been included for the sake of
completeness and the reader familiar with neuroscience can skip it.
In Section \ref{sect:problem} we expose the problem and present the
statistical model, essentially a discretization of the well-known
Morris-Lecar model, and we analyze the model inaccuracies as well.
Next, in Section \ref{sect:methods} we present the different
inference algorithms we apply depending on the knowledge of the
system. Results are given in Section
\ref{sect:seq_estimation:results}, where we tackle three inference
problems: $i)$ when the parameters defining the model are known;
$ii)$ when the parameters of the model are unknown, and thus they
need to be estimated; and, $iii)$ estimation of synaptic
conductances from voltage traces assuming unknown model parameters.
Finally, Section \ref{sect:conclusions} concludes the paper with
final remarks.

\section{Neuron electrophysiology in a nutshell}\label{sect:physiology}

Neurons are the basic information processing structures in the
\ac{CNS} \cite{Dayan05,Izhikevich06,Keener09}. The main function of
a neuron is to receive input information from other neurons, to
process that information, and to send output information to other
neurons. Synapses are connections between neurons, through which
they communicate this information. It is controversial how this
information is encoded, but it is quite accepted that information
produces changes in the electrical activity of the neurons, seen as
voltage changes in the membrane potential (i.e., the difference in
electrical potential between the interior and the exterior of a
biological cell).

The basic constituents of a neuron are the \emph{soma}, which
contains the nucleus of the cell, it is the body of the neuron where
most of the information processing is carried; the \emph{dendrites},
that are extensions of the soma which connect the neuron to
neighboring neurons and capture their stimuli; the \emph{axon},
which is the largest part of a neuron where the information is
transmitted in form of an electrical current. A cell might have only
one axon or more. The physiological meaning for the propagation of
the voltage through the axon can be understood in terms of
voltage-gated ionic channels located in the axon membrane; and the
\emph{synapses}, located at the axon terminal are in charge of the
electrochemical reactions that cause neuron communications to
happen. More precisely, the membrane potential (an electrical
phenomenon) traveling through the axon, when reaching the synapse,
activates the emission of neurotransmitters (a chemical phenomenon)
from the neuron to the receptors of the target neurons. This
chemical reaction is transformed again into electrical impulses in
the dendrites of the receiving neurons.

We are specially interested in understanding the phenomena through
which an electrical voltage travels the axon from the soma to the
synapse. The basic idea is that the membrane covering the axon is
essentially impermeable to most charged molecules. This makes the
axon to act as a capacitor (in terms of electrical circuits) that
separates the inner and outer parts of the neuron's axon. This is
combined with the so-called ionic-channels, that allow the exchange
of intracellular/extracellular ions through electrochemical
gradients. This exchange of ions is responsible for the generation
of an electrical pulse called action potential, that travels along
the neuron's axon. Ionic-channels are found throughout the axon and
are typically voltage-dependent, which is primarily how the action
potential propagates.

The most common ionic species involved in the generation of the
action potential are sodium ($\textrm{Na}^+$), potassium
($\textrm{K}^+$), chloride ($\textrm{Cl}^-$), and calcium
($\textrm{Ca}^{2+}$). For each ionic species, the corresponding
ionic-channel aims at balancing the concentration and electrical
potential gradients, which are opposite forces regulating the
exchange of ions through the gate. The point at which both forces
counterbalance is known as \emph{Nernst equilibrium potential}.

For the sake of simplicity
and without loss of generality, we consider the evolution of the
membrane potential at a specific site of the axon. Therefore,
$v\triangleq v(t)$ denotes the continuous-time membrane potential at
a point in the axon. Accounting that the membrane potential is seen
as a capacitor, the current-voltage relation allows us to express
the total current flowing in the membrane as proportional to the
time derivative of the voltage. Then, we obtain the so-called
\emph{conductance-based} mathematical model for the evolution of
membrane potentials,
\begin{equation}\label{eq:basic_membr_pot}
  C_m \dot{v} = - \sum_{i\in\mathcal{I}} I_i - \bar{g}_L (v - E_L) - I_{\mathrm{syn}} + I_{\mathrm{app}}
\end{equation}
\noindent where $C_m$ is the membrane capacitance and
$I_{\mathrm{app}}$ represents the externally applied currents, for
instance injected via an electrode and used to perform a controlled
experiment.

The time-varying ionic current for the $i$-th ionic species, $i \in
\mathcal{I} = \{\mathrm{Na}, \mathrm{K}, \mathrm{Cl}, \mathrm{Ca},
\dots\}$, is $I_i = \bar{g}_i ~ p_i (v - E_i)$, where $\bar{g}_i$ is
a constant called the maximal conductance, which is fixed for each
ionic species. The variable $p_i \triangleq p_i(v)$ is the average
proportion of channels of the type $i \in \mathcal{I}$ in the open
state. Notice that the proportion is in general voltage-dependent
(i.e., sensitive to the membrane potential) and thus they are said
to be voltage-gated. $p_i$ can be further classified into gates that
activate (i.e., gates that open the $i$-th ionic channel) and those
that inactivate (i.e., gates that close the $i$-th ionic channel).
Mathematically, omitting the dependence on $i$, $p = m^{a} h^{b}$,
where $a$ is the number of activation gates and $0<m\triangleq
m_i(v)<1$ is the probability of activating gate being in the open
state. Similarly, $b$ is the number of inactivation gates and
$0<h\triangleq h_i(v)<1$ the probability of inactivating gate being
in the open state. We refer to $m$ and $h$ as the gating variables
of the ionic channel. The dynamics of these gating variables are
responsible for the membrane potential generation

The \emph{leakage term} in (\ref{eq:basic_membr_pot}), $\bar{g}_L (v
- E_L)$, is mathematically used to gather all ionic channels that
are not explicitly modeled. The maximal conductance of the leakage,
$\bar{g}_L$, is considered constant and it is adjusted to match the
membrane potential at resting state. Similarly, the equilibrium
potential $E_L$ has to be estimated at rest.

$I_{\mathrm{syn}}$ gathers the contribution of neighboring neurons
and it is referred to as the synaptic current. The most general
model for $I_{\mathrm{syn}}$ considers decomposition in 2
independent components:
\begin{equation}
  I_{\mathrm{syn}} = g_\mathrm{E}(t) (v(t) - E_\mathrm{E}) + g_\mathrm{I}(t) (v(t) - E_\mathrm{I})
\end{equation}
\noindent corresponding to excitatory (the most common being
\ac{AMPA} neuroreceptors) and inhibitory (the most common being
\ac{GABA} neuroreceptors) terms, respectively. Roughly speaking,
whereas the excitatory synaptic term makes the postsynaptic neuron
more likely to generate a spike, the inhibitory term makes the
postsynaptic neuron less likely to generate an action potential.
$E_\mathrm{E}$ and $E_\mathrm{I}$ are the corresponding reverse
potentials, typically close to $0$ mV and $-80$ mV, respectively. A
longstanding problem is to characterize the time-varying global
excitatory and inhibitory conductances $g_\mathrm{E}(t)$ and
$g_\mathrm{I}(t)$. There are various mathematical models for the
synaptic current that could be considered
\cite{ErmentroutTerman2010}, here we use the so-called
\emph{effective point-conductance} model of
\cite{Rudolph03,Rudolph2004}. In this model, the
excitatory/inhibitory global conductances are treated as \ac{OU}
processes
    \begin{equation}\label{eq:OUprocess}
      \dot{g}_u(t) = - \frac{1}{\tau_u} (g_u(t) - g_{u,0}) + \sqrt{\frac{2 \sigma_u^2}{\tau_u}} \chi(t)
    \end{equation}
\noindent where $u=\{\mathrm{E},\mathrm{I}\}$. $\chi(t)$ is a
zero-mean, white noise, Gaussian process with unit variance. Then,
the \ac{OU} process has mean $g_{u,0}$, standard deviation
$\sigma_u$, and time constant $\tau_u$. This simple model was shown
in \cite{Rudolph03} to yield a valid description of the synaptic
noise, capturing the properties of more complex models.

Conductance-based models like (\ref{eq:basic_membr_pot}) are widely
used, mostly varying on the number and type of gating variables and
the activation/inactivation functions defining the dynamics of the
gating variables. The pioneer work by Hodgkin and Huxley in
\cite{Hodgkin52} has been followed by a plethora of alternative
models such as
\cite{FitzHugh61,Nagumo62,Morris81,Izhikevich04,Rabinovich06} to
name a few.

\begin{figure}[t]
 \begin{center}
  \includegraphics[width=\columnwidth]{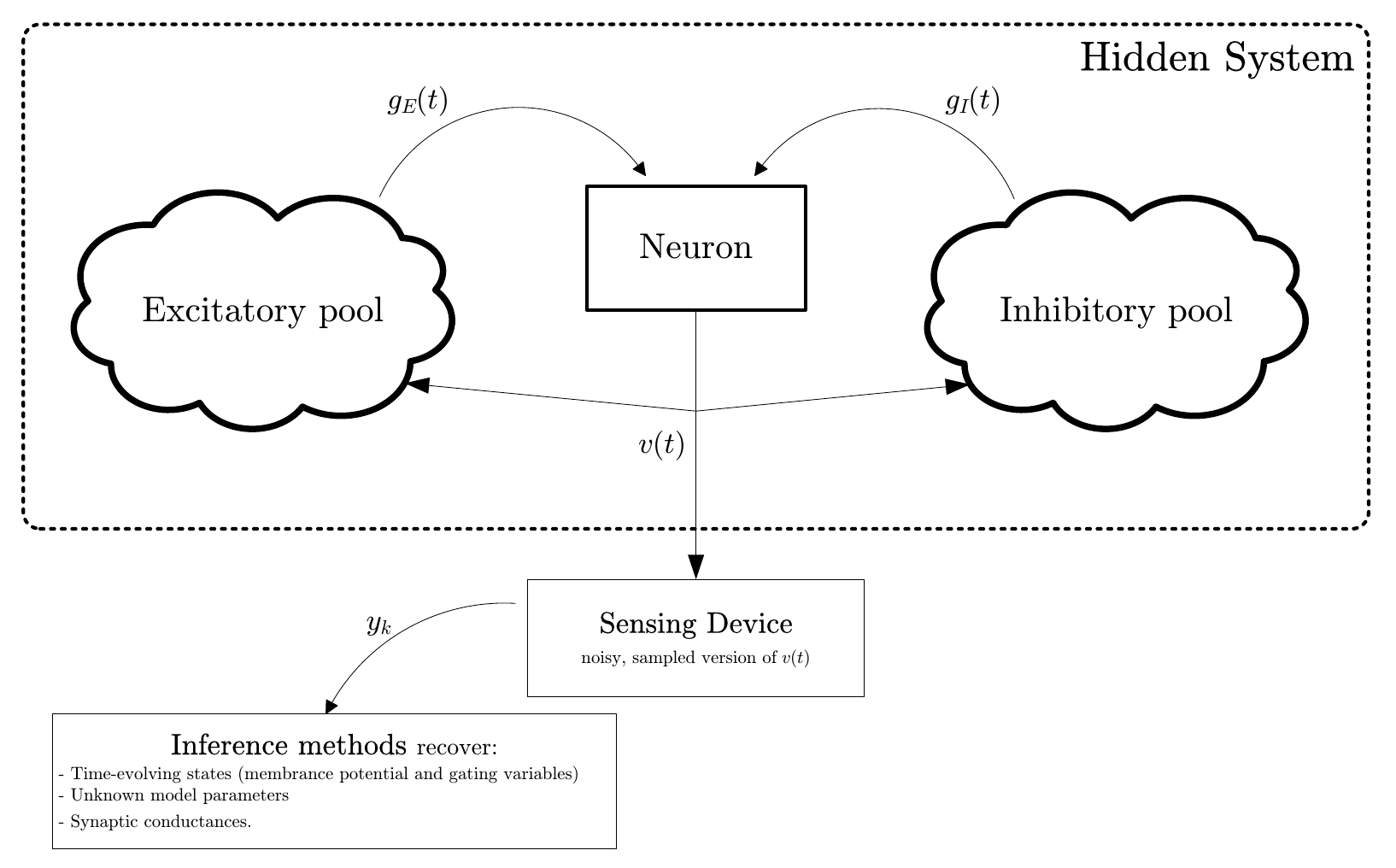}
  \caption{The experimental setup of interest in this paper.}\label{fig:synaptic_cond}
 \end{center}
\end{figure}

Fig. \ref{fig:synaptic_cond} shows the basic setup we are dealing
with in this article. The neuron under observation has its own
dynamics, producing electrical voltage patterns. The generation of
action potentials is regulated by internal drivers (e.g., the active
gating variables of the neuron) as well as exogenous factors like
excitatory and inhibitory synaptic conductances produced by pools of
connected neurons. This system is unobservable, in the sense that we
cannot measure it directly. The sole observation from this system
are the noisy membrane potentials $y_k$. In this experimental
scenario, we aim at applying sequential inference methods to extract
the following quantities:
\begin{enumerate}
  \item The time-evolving states characterizing the neuron dynamics,
  including a filtered membrane potential and the dynamics of the gating variables.
  \item The parameters defining the neuron model. It was seen that
  conductance-based models require the knowledge (or adjustment) of
  a number of static parameters. It is desirable to have autonomous
  inference algorithms that estimate these parameters as well, on
  the top of the time-evolving states.
  \item The dynamics of synaptic conductances and its parameters, the final goal being to discern the temporal contributions of global excitation from those of global
  inhibition, $g_\mathrm{E}(t)$ and $g_\mathrm{I}(t)$ respectively.
\end{enumerate}

\section{Problem statement and model}\label{sect:problem}

The membrane potential, obtained from intracellular recordings, is
one of the most valuable signals of neurons' activity. Most of the
neuron models have been derived from fine measurements and allow the
progress of ``in silico'' experiments. The recording of the membrane
potential is a physical process, including some
approximations/inaccuracies involving:
\begin{enumerate}
  \item Voltage observations are noisy. This is due, in part, to the thermal
  noise at the sensing device, non-ideal conditions in experimental
  setups, etc.
  \item Recorded observations are discrete. All sensing devices record data
  by sampling at regular time intervals the continuous-time
  natural phenomena. This is the task of an \ac{ADC}. Moreover, these samples are typically
  digitized, i.e. expressed by a finite number of bits. This latter
  issue is not tackled in the work as we assume that modern
  computer capabilities allow us to sample with relatively large
  number of bits per sample.
\end{enumerate}

The problem investigated in this paper considers recordings of noisy
voltage traces to infer the hidden gating variables of the neuron
model, as well as filtered voltage estimates, model parameters and
input synaptic conductances. Data is recorded at discrete
time-instants $k$ at a sampling frequency $f_s=1/ \mathrm{T}_s$. The
problem can thus be posed in the form of a discrete-time,
state-space model. The observations are
\begin{equation}
  y_k \sim \mathcal{N} (v_k,\sigma_{y,k}^2) ~,
\end{equation}
\noindent with $v_k$ representing the nominal membrane potential and
$\sigma_{y,k}^2$ modeling the noise variance due to the sensor or
the instrumentation inaccuracies when performing the experiment. To
provide comparable results, we define the signal-to-noise ratio
($\mathrm{SNR}$) as $\mathrm{SNR}=P_s/P_n$, with $P_s$ being the
average signal power and $P_n=\sigma_{y,k}^2$ the noise power.

The methods presented in this paper rely on models for the evolution
of the voltage-traces and the hidden variables of a neuron. For
instance, the conductance-based neuron models and the \ac{OU}
processes for the synaptic conductances introduced in Section
\ref{sect:physiology}.

Concerning neuron models, in this work, we focus on the Morris-Lecar
model \cite{Morris81} for the sake of clarity. This model is very
popular in computational neuroscience as it is able to model a wide
variety of neural dynamics. Details of the Morris-Lecar model can be
consulted in Appendix \ref{ap:MLmodel}. The unknown state vector in
this case is composed of the membrane potential and the K$^+$ gating
variable,
\begin{equation}
\bfx_k = \left(\begin{array}{c}
                       v_k \\
                       n_k
                     \end{array} \right) ~.
\end{equation}

Notice that the Morris-Lecar neuron model is defined by a system of
continuous-time, \acp{ODE} of the form $\dot{\bfx}= f(\bfx)$. In
general, mathematical models for neurons are of this type. However,
due to the sampled recording of measurements, we are interested in
expressing the model in the form of a discrete state-space,
\begin{equation}\label{eq:MarkovianForm}
  \bfx_k = f_k (\bfx_{k-1}) + \bm{\nu}_k ~,
\end{equation}
\noindent where $\bm{\nu}_k \sim
\mathcal{N}(\mathbf{0},\bm{\Sigma}_{x,k})$ is the process noise
which includes the model inaccuracies. The covariance matrix
$\bm{\Sigma}_{x,k}$ is used to quantify our confidence in the model
that maps $f_k: \{v_{k-1} , n_{k-1}\} \mapsto \{v_{k} , n_{k}\}$. In
general, obtaining a closed-form analytical expression for $f_k$
without approximations is not possible.
In such a case, we could use the Euler method:
\begin{equation}
  \dot{\bfx} \doteq  \frac{d\bfx}{dt} \approx \frac{\Delta \bfx}{\Delta t} = \frac{\bfx(t+\Ts) - \bfx(t)}{\Ts} = f(\bfx(t)),
\end{equation}
\noindent where $\Delta t=\Ts$ is the sampling period. Thus, we can
write (\ref{eq:MarkovianForm}) as
\begin{equation}
  \bfx_k = \bfx_{k-1} + \Ts f(\bfx_{k-1}) ~,
\end{equation}
\noindent which is of the Markovian type.

If we focus on the Morris-Lecar model, the resulting discrete
version of the \ac{ODE} system in
(\ref{eq:ML_ODE1})-(\ref{eq:ML_ODE2}) is:
\begin{eqnarray}
\nonumber v_k &=& v_{k-1} - \frac{\Ts}{C_m} \Big( \bar{g}_\mathrm{L} (v_{k-1}-E_\mathrm{L}) \\
\nonumber {}&+& \bar{g}_\mathrm{Ca} m_\infty(v_{k-1}) (v_{k-1}-E_\mathrm{Ca})\\
{}&+& \bar{g}_\mathrm{K} n_{k-1} (v_{k-1}-E_\mathrm{K}) - I_{\mathrm{app}} \Big) \label{eq:ML_discrete1} \\
n_k &=& n_{k-1} + \Ts \phi \frac{n_\infty(v_{k-1})-n_{k-1}}{\tau_{n}(v_{k-1})}
\label{eq:ML_discrete2} ~,
\end{eqnarray}
\noindent with $m_\infty(v_k)$, $n_\infty(v_k)$, and $\tau_n(v_k)$
defined as in (\ref{eq:sigmoids1})-(\ref{eq:sigmoids3}), see
Appendix \ref{ap:MLmodel}. Then, (\ref{eq:ML_discrete1}) and
(\ref{eq:ML_discrete2}) can be interpreted as $\bfx_k = f_k
(\bfx_{k-1})$.

The goal is to express the inference problem in state-space
formulation and apply stochastic filtering tools learned from
signal processing. The final ingredient to do so is to introduce the
so-called process noise in the state equation
\begin{equation}\label{eq:seq_est:genmodel}
  \bfx_k = f_k (\bfx_{k-1}) + \left( \begin{array}{c}
                                \nu_{v,k}  \\
                                \nu_{n,k}
                              \end{array} \right)  ~,
\end{equation}
\noindent where the noise terms $\nu_{v,k}$ and $\nu_{n,k}$ are
assumed jointly Gaussian with covariance matrix $\bm{\Sigma}_{x,k}$.
Further details of this matrix are discussed in Section
\ref{sect:seq_estimation:inaccuracies}.

This general form of Markovian type is preserved when the model is
extended with a couple of \ac{OU} processes associated to the
excitatory and inhibitory synaptic conductances. In this case, the
resulting state-space model is composed by the Morris-Lecar model
used so far (with the peculiarity that the term $-I_{\mathrm{syn}}$
is added to (\ref{eq:ML_ODE1})), plus the \ac{OU} stochastic process
in (\ref{eq:OUprocess}) describing $I_{\mathrm{syn}}$. Therefore,
the continuous-time state is
$\bfx=(v,n,g_\mathrm{E},g_\mathrm{I})^\top$. The discrete version of
the state-space is used again.

\subsection{Model inaccuracies} \label{sect:seq_estimation:inaccuracies}

The proposed estimation method relies on the fact that the neuron
model is known. This is true to some extent, but most of the
parameters in the Morris-Lecar model discussed are to be estimated.
Typically this model calibration is done beforehand, but as we will
see later in Section \ref{subsect:seq_estimation:method2} this could
be done in parallel to the filtering process. Therefore, the
robustness of the method to possible inaccuracies should be
assessed. In this section, we point out possible causes of
mismodeling. Computer simulations are later used to characterize the
performance of the proposed methods under these impairments.

In the single-neuron model considered, three major sources of
inaccuracies can be identified. First, the applied current
$I_{\mathrm{app}}$ can be itself noisy, with a variance depending on
the quality of the instrumentation used and the experiment itself.
We model the actual applied current as a random variable
\begin{equation}\label{eq:Iapp}
  I_{\mathrm{app}} = I_o + \nu_{I,k} ~,~ \nu_{I,k} \sim \mathcal{N} (0 , \sigma^2_{I}) ~,
\end{equation}
\noindent where $I_o$ is the nominal current applied and
$\sigma^2_{I}$ the variance around this value. Plugging
(\ref{eq:Iapp}) into (\ref{eq:ML_discrete1}) we obtain that the
contribution of $I_{\mathrm{app}}$ to the noise term is
$\frac{\Ts}{C_m}\nu_{I,k} \sim \mathcal{N} (0 ,
(\Ts/C_m)^2\sigma^2_{I})$.

Secondly, when the conductance of the leakage term is estimated
beforehand, some inaccuracies might be taken into account. In
general, this term is considered constant in the models although it
gathers relatively distinct phenomena that can potentially be
time-varying. The maximal conductance of the leakage term is
therefore inaccurate and modeled as
\begin{equation}\label{eq:gl}
  \bar{g}_L = \bar{g}_L^o + \nu_{g,k} ~,~ \nu_{g,k} \sim \mathcal{N} (0 , \sigma^2_{g})~,
\end{equation}
\noindent where $\bar{g}_L^o$ is the nominal, estimated conductance
and $\sigma^2_{g}$ the variance of this estimate. Similarly,
inserting (\ref{eq:gl}) into (\ref{eq:ML_discrete1}) we see that the
contribution of $\bar{g}_L$ to the noise term is
$\frac{\Ts}{C_m}\nu_{g,k} \sim \mathcal{N} (0 ,
(\Ts/C_m)^2(v_{k-1}-E_\mathrm{L})\sigma^2_g)$.

Finally, the parameters in $m_\infty(v_k)$, $n_\infty(v_k)$, and
$\tau_n(v_k)$ are to be estimated. In general, these parameters can
be properly obtained off-line by standard methods, see
\cite{Izhikevich06}. However, as they are estimates, a residual
error typically remains. To account for these inaccuracies, we
consider that the equation governing the evolution of gating
variables is corrupted by a zero-mean additive white Gaussian
process with variance $\sigma_n^2$.

This analysis allows us to construct a realistic
$\bm{\Sigma}_{x,k}$, as the contribution of the aforementioned
inaccuracies. In a practical setup, in order to compute the noise
variance due to leakage, we need to use the approximation
$\hat{v}_{k-1} \approx v_{k-1}$, where $\hat{v}_{k-1}$ is estimated
by the filtering method. We construct the covariance matrix of the
model as
\begin{equation}
  \bm{\Sigma}_{x,k} = \left( \begin{array}{cc}
                               \sigma_v^2 & 0 \\
                               0 & \sigma_n^2
                             \end{array} \right)   ~,
\end{equation}
\noindent where we used that the overall noise in the voltage model
is $\frac{\Ts}{C_m}(\nu_{I,k} - \nu_{g,k}) \sim \mathcal{N} (0 ,
\sigma^2_v)$ and
\begin{equation}
\sigma^2_v = \left(\frac{\Ts}{C_m}\right)^2 \left(\sigma^2_{I} +
(\hat{v}_{k-1}-E_\mathrm{L})^2 \sigma^2_{g}\right)
\end{equation}
\noindent as an estimate of $\sigma_v^2$, provided accurate
knowledge of $\sigma^2_{I}$ and $\sigma^2_{g}$. Otherwise, the
covariance matrix of the process could be estimated by other means,
as the ones presented in Section
\ref{subsect:seq_estimation:method2} for mixed state-parameter
estimation in nonlinear filtering problems.

\section{Methods}\label{sect:methods}

Two filtering methods are proposed here, depending on the knowledge
regarding the underlying dynamical model. Section
\ref{subsect:seq_estimation:method} presents an algorithm able to
estimate the states in $\bfx_k$ by a \ac{PF} methodology, the
particularity being that an optimal distribution is used to draw the
random samples characterizing the joint filtering distribution of
interest. This method assumes knowledge of the parameter values of
the system model, although we account for some inaccuracies as
detailed in Section \ref{sect:seq_estimation:inaccuracies}. An
enhanced version of this method is presented in Section
\ref{subsect:seq_estimation:method2}, where we relax the assumption
of knowing the parameter values. Leveraging on a \ac{PMCMC}
algorithm and the use of the optimal importance density as in the
first method, we present a method that is able to filter $\bfx_k$
while estimating the values describing the neuron model.

\subsection{Sequential estimation of voltage traces and gating variables}\label{subsect:seq_estimation:method}

The type of problems we are interested in involve the estimation of
time-evolving signals that can be expressed through a state-space
formulation. Particularly, estimation of the states in a
single-neuron model from noisy voltage traces can be readily seen as
a filtering problem. Bayesian theory provides the mathematical tools
to deal with such problems in a systematic manner. The focus is on
sequential methods that can incorporate new available measurements
as they are recorded without the need for reprocessing all past
data.

Bayesian filtering involves the recursive estimation of states
$\bfx_k \in {\mathbb R}^{n_x}$ given measurements $y_k \in {\mathbb
R}$ at time $t$ based on all available measurements, ${y}_{1:k} =
\left\lbrace {y}_{1},\dots , {y}_k \right\rbrace $. To that aim, we
are interested in the filtering distribution $p({ \bfx}_k|{
y}_{1:k})$, which can be recursively expressed as
\begin{equation}\label{eq:posterior_recursion}
p({ \bfx}_k | { y}_{1:k}) =  \frac{p({ y}_k | { \bfx}_k) p({
\bfx}_k| { \bfx}_{k-1})}{p({ y}_k|{ y}_{1:k-1})} p({ \bfx}_{k-1} | {
y}_{1:k-1}) ~,
\end{equation}
\noindent with $p({ y}_k | { \bfx}_k)$ and $ p({ \bfx}_k| {
\bfx}_{k-1})$ referred to as the likelihood
and the prior distributions, respectively. 
Unfortunately, (\ref{eq:posterior_recursion}) can only be obtained
in closed-form in some special cases and in more general setups we
should resort to more sophisticated methods. In this paper we
consider PF to cope with the nonlinearity of the model. Although
other lighter approaches might be possible as well \cite{Chen03}, we
seek the maximum accuracy regardless the involved computational
cost. Theoretically, for sufficiently large number of particles,
particle filters offer such performances.

Particle filters, see
\cite{SeqMC01,Arulampalam02,DjuricM03,BeyondKF}, approximate the
filtering distribution by a set of $N$ weighted random samples,
forming the random measure $\left\{{ \bfx}^{(i)}_k,w^{(i)}_k
\right\}_{i=1}^{N}$. These random samples are drawn from the
importance density distribution, $\pi(\cdot)$,
\begin{equation}\label{eq:gen_importance_fun}
{ \bfx}^{(i)}_k \sim \pi({ \bfx}_k|{ \bfx}^{(i)}_{0:k-1}, {
y}_{1:k})
\end{equation}
\noindent and weighted according to the general formulation
\begin{equation}\label{eq:gen_importance_w}
w^{(i)}_k \propto w^{(i)}_{k-1}\frac{p({ y}_k | { \bfx}_{0:k}^{(i)}
, { y}_{1:k-1}) p({ \bfx}_k^{(i)} | { \bfx}_{k-1}^{(i)})}{\pi({
\bfx}_k^{(i)} | { \bfx}_{0:k-1}^{(i)},{ y}_{1:k} )}  ~.
\end{equation}

The importance density from which particles are drawn is a key issue
in designing efficient PFs. It is well-known that the optimal
importance density is $\pi({ \bfx}_k|{ \bfx}^{(i)}_{0:k-1}, {
y}_{1:k}) = p({ \bfx}_k|{ \bfx}^{(i)}_{k-1}, { y}_{k}),$ in the
sense that it minimizes the variance of importance weights. Weights
are then computed using (\ref{eq:gen_importance_w}) as $w^{(i)}_k
\propto w^{(i)}_{k-1} p({ y}_k|{ \bfx}^{(i)}_{k-1})$. This choice
requires the ability to draw from $p({ \bfx}_k|{ \bfx}^{(i)}_{k-1},
{ y}_{k})$ and to evaluate $p({ y}_k|{ \bfx}^{(i)}_{k-1})$. In
general, the two requirements cannot be met and one needs to resort
to suboptimal choices. However, we are able to use the optimal
importance density since the state-space model assumed here is
Gaussian, with nonlinear process equations but related linearly to
observations \cite{Doucet00}. The equations are:
\begin{equation}
  p({ \bfx}_k|{ \bfx}^{(i)}_{k-1}, { y}_{k}) = \mathcal{N}(\bm{\mu}_{\pi,k}^{(i)},\bm{\Sigma}_{\pi,k})
\end{equation}
\noindent with
\begin{eqnarray}
  \bm{\mu}_{\pi,k}^{(i)} &=& \bm{\Sigma}_{\pi,k} \left( \bm{\Sigma}_{x,k}^{-1} f_k({ \bfx}^{(i)}_{k-1}) + \frac{y_k}{\sigma_{y,k}^2} \right) \\
  \bm{\Sigma}_{\pi,k} &=& \left( \bm{\Sigma}_{x,k}^{-1} + \sigma_{y,k}^{-2} \mathbf{I} \right)^{-1} ~,
\end{eqnarray}
\noindent and the importance weights can be updated using
\begin{equation}
  p({ y}_k|{ \bfx}^{(i)}_{k-1}) = \mathcal{N}(\mathbf{h}^\top f_k({ \bfx}^{(i)}_{k-1}), \mathbf{h}^\top \bm{\Sigma}_{x,k} \mathbf{h} + \sigma_{y,k}^2)
  ~,
\end{equation}
\noindent with $\mathbf{h} = (1,0)^\top$. The PF provides a discrete
approximation of the filtering distribution of the form $p({
\bfx}_k|{ y}_{1:k}) \approx \sum_{i=1}^N w^{(i)}_k
\delta(\mathbf{\bfx}_k - \bfx^{(i)}_k)$, which gather all
information from $\bfx_{k}$ that the measurements up to time $k$
provide. The minimum mean square error estimator can be obtained as
\begin{equation}\label{eq:MMSEpf2}
    \hat{\bfx}_{k} = \sum_{i=1}^N w^{(i)}_k \bfx^{(i)}_{k}
    ~,
\end{equation}
\noindent where $\hat{\bfx}_k = \left(\hat{v}_k , \hat{n}_k
\right)^\top$. Recall that the method discussed in this section
could be easily adapted to other neuron models by simply
substituting the corresponding transition function $f_k$ and
constructing the state vector $\bfx_k$ accordingly.

As a final step, PFs incorporate a resampling strategy to avoid
collapse of particles into a single state point. Resampling consists
in eliminating particles with low importance weights and replicating
those in high-probability regions \cite{Douc05}. The overall
algorithm can be consulted in Algorithm \ref{alg:optSPF} at instance
$k$. Notice that this version of the algorithm requires knowledge of
noise statistics and all the model parameters, which for the
Morris-Lecar model are
\begin{equation}\label{eq:ML_modelparams}
    \bm{\Theta}=(\bar{g}_\mathrm{L}, E_\mathrm{L}, \bar{g}_\mathrm{Ca}, E_\mathrm{Ca}, \bar{g}_\mathrm{K}, E_\mathrm{K}, \phi, V_1, V_2, V_3, V_4 )^\top ~.
\end{equation}

When we add the dynamics of the synaptic conductances, the vector
$\bm{\Theta}$ of model parameters also includes $\tau_E$, $\tau_I$,
$g_{E,0}$, $g_{I,0}$, $\sigma_{E}$ and $\sigma_{I}$.

\begin{algorithm}
\caption{Particle filtering with optimal importance density}
\label{alg:optSPF}
\begin{algorithmic}[1]
\REQUIRE $\bm{\Sigma}_{x,k}$, $\sigma_{y,k}^2$, $\bm{\Theta}$, $ \left\{\bfx^{(i)}_{k-1},w^{(i)}_{k-1} \right\}_{i=1}^{N}$ and $y_k$
\ENSURE $\left\{\bfx^{(i)}_k,w^{(i)}_k \right\}_{i=1}^{N} $ and $\hat{\bfx}_k$
\STATE Calculate $\bm{\Sigma}_{\pi,k} = \left( \bm{\Sigma}_{x,k}^{-1} + \sigma_{y,k}^{-2} \mathbf{I} \right)^{-1}$
\FOR{$i=1$ to $N$}
\STATE ~~~~ Calculate $\bm{\mu}_{\pi,k}^{(i)} = \bm{\Sigma}_{\pi,k} \left( \bm{\Sigma}_{x,k}^{-1} f_k({ \bfx}^{(i)}_{k-1}) + \frac{y_k}{\sigma_{y,k}^2} \right)$
\STATE ~~~~ Generate $\bfx^{(i)}_k \sim \mathcal{N}(\bm{\mu}_{\pi,k}^{(i)},\bm{\Sigma}_{\pi,k})$
\STATE ~~~~ Calculate $\tilde{w}^{(i)}_k = w^{(i)}_{k-1}\frac{p(y_k | {\bfx}_{0:k}^{(i)} , y_{1:k-1}) p(\bfx_k^{(i)} | {\bfx}_{k-1}^{(i)})}{\mathcal{N}(\bm{\mu}_{\pi,k}^{(i)},\bm{\Sigma}_{\pi,k})} $
\ENDFOR
\FOR{$i=1$ to $N$}
\STATE ~~~~ Normalize weights: $w^{(i)}_k=\frac{\tilde{w}^{(i)}_k}{\sum_{j=1}^{ N}\tilde{w}^{(j)}_k}$
\ENDFOR
\STATE MMSE state estimation: $\hat{\bfx}_k = \sum_{i=1}^N w^{(i)}_k {\bfx}^{(i)}_k$
\STATE $\left\{\bfx^{(i)}_k,1/N \right\}_{i=1}^{N}$ = Resample($\left\{\bfx^{(i)}_k, \hspace{0.1cm} w^{(i)}_k \right\}_{i=1}^{N}$)
\end{algorithmic}
\end{algorithm}

\subsection{Joint estimation of states and model parameters}\label{subsect:seq_estimation:method2}

In practice the parameters in (\ref{eq:ML_modelparams}) might not be
known. It is reasonable to assume that $\bm{\Theta}$, or a subset of
these parameters $\btheta \subseteq \bm{\Theta}$, are unknown and
need to be estimated at the same time the filtering method in
Algorithm \ref{alg:optSPF} is executed. Therefore, the ultimate goal
in this case is to estimate jointly the time evolving states and the
unknown parameters of the model, $\bfx_{1:k}$ and $\btheta$
respectively.

Joint estimation of states and parameters is a longstanding problem
in Bayesian filtering, and specially hard to handle in the context
of \acp{PF}. Refer to \cite{Andrieu04,Andrieu05,Poyiadjis11} and
their references for a complete survey. Here, we follow the approach
in \cite{Andrieu10} and make us of the so-called \ac{PMCMC} to
enhance the presented \ac{PF} algorithm with parameter estimation
capabilities. In the remainder of this section we provide the basic
ideas to use the algorithm, following a similar approach as in
\cite{Sarkka13book}.

Following the Bayesian philosophy we adopt here, the problem
fundamentally reduces to assigning an a priori distribution for the
unknown parameter $\btheta \in \mathbb{R}^{n_\theta}$ and extending
the state-space model (here we adopt its probabilistic
representation)
\begin{eqnarray}
    \bm{\theta}  & \sim & p(\btheta) \\
    \bfx_k &\sim & p(\bfx_{k} | \bfx_{k-1}, \btheta) \quad \textrm{for } k\geq 1 \\
    y_k &\sim & p(y_{k} | \bfx_{k}, \btheta) \quad \textrm{for } k\geq 1
\end{eqnarray}
\noindent and initial state distribution $\bfx_0 \sim p(\bfx_{0} |
\btheta)$. Applying Bayes' rule, the full posterior distribution can
be expressed as
\begin{equation}
  p(\bfx_{0:T} , \btheta | y_{1:T}) = \frac{p(y_{1:T} | \bfx_{0:T} , \btheta) p(\bfx_{0:T} | \btheta) p(\btheta)}{p(y_{1:T})}
\end{equation}
\noindent with
\begin{eqnarray}
  p(y_{1:T} | \bfx_{0:T} , \btheta) &=& \prod_{k=1}^T p(y_{k} | \bfx_{k}, \btheta) \\
  p(\bfx_{0:T} | \btheta) &=& p(\bfx_0 | \btheta) \prod_{k=1}^T p(\bfx_{k} | \bfx_{k-1}, \btheta) ~.
\end{eqnarray}
Notice here that we are dealing with a finite horizon of
observations $T$. Then, from a Bayesian perspective, the estimation
of $\btheta$ is equivalent to obtaining its marginal posterior
distribution $p(\btheta | y_{1:T}) = \int p(\bfx_{0:T} , \btheta |
y_{1:T}) d\bfx_{0:T}$. However, this is in general extremely hard to
compute analytically and one needs to find workarounds. Evaluation
of the full posterior turns to be not only computationally
prohibitive, but useless if states cannot be marginalized out
analytically. Alternative methods resort on the factorization of the
parameter marginal distribution as $p(\btheta | y_{1:T}) = p(y_{1:T}
| \btheta) p(\btheta)$ and how Bayesian filters can be transformed
to provide characterizations of the marginal likelihood distribution
$p(y_{1:T} | \btheta)$ and related quantities. The marginal
likelihood distribution can be recursively factorized in terms of
the predictive distributions of the observations: $p(y_{1:T} |
\btheta) = \prod_{k=1}^T p(y_{k} | y_{1:k-1}, \btheta)$, with
$p(y_{k} | y_{1:k-1}, \btheta) = \int p(y_{k} | \bfx_{k}, \btheta)
p(\bfx_{k} | y_{1:k-1} , \btheta) d\bfx_{k}$ obtained
straightforwardly as a byproduct of any of the Bayesian filtering
methods, see \cite{Chen03}.

A useful transformation of the marginal likelihood is the so-called
\emph{energy function}, which is sometimes more convenient to deal
with. The energy function is defined as $\varphi_T(\btheta) = - \ln
p(y_{1:T} | \btheta) - \ln p(\btheta)$ or, equivalently, $p(\btheta
| y_{1:T}) \propto \exp(- \varphi_T(\btheta))$. The energy function
can then be recursively computed as a function of the predictive
distribution
\begin{eqnarray}
  \varphi_0(\btheta) &=& - \ln p(\btheta) \\
  \varphi_k(\btheta) &=& \varphi_{k-1}(\btheta) - \ln p(y_{k} | y_{1:k-1}, \btheta) ~ \textrm{for } k\geq 1
\end{eqnarray}

Then, the basic problem is to obtain an estimate of the predictive
distribution $p(y_{k} | y_{1:k-1}, \btheta)$ from the \ac{PF} we
have designed in Section \ref{subsect:seq_estimation:method} and use
it in conjunction with $p(\btheta)$ to infer the marginal
distribution $p(\btheta | y_{1:T})$ of interest. This latter step
can be performed in several ways, from which we choose to use the
\ac{MCMC} methodology to continue with a fully Bayesian solution.
Besides, it is known to be the solution that provides best results
when used in a \ac{PF}. Next, we detail how $\varphi_k(\btheta)$ can
be obtained from a \ac{PF} algorithm, we present the \ac{MCMC}
method for parameter estimation, and finally we sketch the overall
algorithm.

\subsubsection{Computing the energy function from particle filters}
The modification needed is rather small. Actually, it is
non-invasive in the sense that the algorithm remains the same and
the energy function can be computed adding some extra formulae.
Recall that the predictive distribution $p(y_{k} | y_{1:k-1},
\btheta)$ is composed of two distributions and that the \ac{PF}
provides characterizations of these two distributions. Then, one
could use a particle approximation $p(y_{k} | y_{1:k-1}, \btheta)
\approx \hat{p}(y_{k} | y_{1:k-1}, \btheta) = \sum_{i=1}^N
w^{(i)}_{k-1} \zeta^{(i)}_{k}$ with $w^{(i)}_{k-1}$ as in the
original algorithm and
\begin{equation}
    \zeta^{(i)}_{k} = \frac{p(y_{k} | \bfx_{k}^{(i)},\btheta) p(\bfx_{k}^{(i)} | \bfx_{k-1}^{(i)},\btheta)}{\pi
(\bfx_{k}^{(i)} | \bfx_{0:k-1}^{(i)},y_{1:k},\btheta)} ~.
\end{equation}

Then, it is straightforward to identify the energy function approximation as
\begin{eqnarray}
    \varphi_T(\btheta) &\approx& - \ln p(\btheta) - \sum_{k=1}^T \ln \hat{p}(y_{k} | y_{1:k-1}, \btheta) \\
    {}&=& - \ln p(\btheta) - \sum_{k=1}^T \ln \sum_{i=1}^N w^{(i)}_{k-1} \zeta^{(i)}_{k} = \hat{\varphi}_T(\btheta) \label{eq:enfun_est}
\end{eqnarray}
\noindent which can be computed recursively in the \ac{PF} algorithm.

\subsubsection{The Particle Markov-Chain Monte-Carlo algorithm}
Once an approximation of the energy function is available, we can
apply \ac{MCMC} to infer the marginal distribution of the
parameters. \ac{MCMC} methods constitute a general methodology to
generate samples recursively from a given distribution by randomly
simulating from a Markov chain
\cite{Gilks96,Berzuini97,Liu05,Brooks11,Donnet2014}. There are many
algorithms implementing the \ac{MCMC} concept, being one of the most
popular the Metropolis-Hastings (MH) algorithm. At the $j$-th
iteration, the MH algorithm samples a candidate point $\btheta^\ast$
from a proposal distribution $q(\btheta^\ast | \btheta^{(j-1)})$
based on the previous sample $\btheta^{(j-1)}$. Starting from an
arbitrary value $\btheta^{(0)}$, the MH algorithm accepts the new
candidate point (meaning that it was generated from the target
distribution, $p(\btheta | y_{1:T})$) using the rule
\begin{equation}
    \btheta^{(j)} = \left\{ \begin{array}{cc}
                              \btheta^\ast,  & \textrm{if } u\leq \alpha^{(j)}  \\
                              \btheta^{(j)}, & \textrm{otherwise}
                            \end{array}
     \right.
\end{equation}
\noindent where $u$ is drawn randomly from a uniform distribution,
$u \sim \mathcal{U}(0,1)$, and
\begin{equation}
    \nonumber \alpha^{(j)} = \min \left\{ 1, \exp(\varphi_T(\btheta^{(j-1)}) - \varphi_T(\btheta^{\ast})) \frac{q(\btheta^{(j-1)}|\btheta^\ast)}{q(\btheta^\ast | \btheta^{(j-1)})} \right\}
\end{equation}
\noindent is referred to as the acceptance probability.

It is critical for the performance of the algorithm the choice of
the proposal density. A common choice is $q(\btheta | \btheta^{(j-1)}) = \mathcal{N} (\btheta ; \btheta^{(j-1)}, \bm{\Sigma}^{(j-1)})$ with the selection of the transitional covariance
remaining as the tuning $\bm{\Sigma}^{(j-1)}$ parameter. This covariance
can be adapted as iterations of the MCMC method progress. In this
work we have adopted the \ac{RAM} algorithm \cite{Vihola12}. The
\ac{RAM} algorithm can be consulted in Algorithm \ref{alg:RAM}.
We use the notation that ${\mathbf S}=\mathrm{Chol}\left({\mathbf A}\right)$
denotes the Cholesky factorization of an Hermitian positive-definite matrix $\mathbf{A}$ such that ${\mathbf A} = {\mathbf S}{\mathbf S}^\top$, and
${\mathbf S}$ is a lower triangular matrix \cite{Golub96}.
The \ac{RAM} algorithm outputs a set of samples $\left\{\btheta^{(j)}\right\}_{j=1}^{M}$, where $M$ is the number of iterations of the \ac{MCMC} procedure.
This samples are originated from the target distribution $\left\{ \btheta^{(j)} \sim p(\btheta | y_{1:T}) \right\}_{j=1}^{M} ~,$ which can be used to approximate (after neglecting the first samples corresponding to the transient phase of the algorithm) it as
\begin{equation}
    p(\btheta | y_{1:T}) \approx \frac{1}{M} \sum_{j=1}^{M} \delta(\btheta - \btheta^{(j)}) ~,
\end{equation}
\noindent and one can obtain the desired statistics from the characterization of the marginal distribution. For instance, point estimates of the
parameter like
\begin{equation}\label{eq:theta_est}
    \hat{\btheta}^\textrm{MMSE} = \frac{1}{M} \sum_{j=1}^{M} \btheta^{(j)} \qquad \textrm{or} \qquad \hat{\btheta} = \btheta^{(M)}  ~.
\end{equation}

The main assumption in Algorithm \ref{alg:RAM} is the ability of
evaluating the energy function, $\varphi_T(\cdot)$. We have seen
earlier how this can be done in a \ac{PF}. Roughly speaking, the
\ac{PMCMC} algorithm consists on putting together these two
algorithms \cite{Andrieu10}. The resulting \ac{PMCMC} method can be
consulted in Algorithm \ref{alg:optSPF_MCMC}.

\begin{algorithm}[ht]
\caption{Robust Adaptive Metropolis}\label{alg:RAM}
\begin{algorithmic}[1]
\REQUIRE $M$, $\btheta^{(0)}$, $\bm{\Sigma}^{(0)}$, $\gamma\in(\frac{1}{2},1]$, $\bar{\alpha}_\ast$, and $\varphi_T(\cdot)$
\ENSURE $\left\{\btheta^{(j)}\right\}_{j=1}^{M} $
\STATE Initialize: $\mathbf{S}_0 = \mathrm{Chol}\left( \bm{\Sigma}^{(0)} \right)$ and $\varphi_T(\btheta^{(0)})=0$
\FOR{$j=1$ to $M$}
\STATE Draw $\mathbf{a} \sim \mathcal{N} (\mathbf{0}, \mathbf{I} )$
\STATE Compute $\btheta^{\ast} = \btheta^{(j-1)} + \mathbf{S}_{j-1} \mathbf{a} $
\STATE Compute $\alpha^{(j)} = \min \left\{ 1, \exp(\varphi_T(\btheta^{(j-1)}) - \varphi_T(\btheta^{\ast})) \right\}$
\STATE Draw $u \sim \mathcal{U}(0,1)$
\IF{$u\leq \alpha^{(j)}$}
\STATE $\btheta^{(j)} = \btheta^\ast$
\ELSE
\STATE  $\btheta^{(j)} = \btheta^{(j-1)}$
\ENDIF
\STATE Compute $\eta^{(j)} = j^{-\gamma}$
\STATE Compute $\mathbf{D}_{j} = \left( \mathbf{I} + \eta^{(j)} (\alpha^{(j)} - \bar{\alpha}_\ast) \frac{\mathbf{a}\mathbf{a}^\top}{||\mathbf{a}||^2}\right)$
\STATE Compute $\mathbf{S}_{j} = \mathrm{Chol} \left( \mathbf{S}_{j-1} \mathbf{D}_{j} \mathbf{S}_{j-1}^\top \right)$
\ENDFOR
\end{algorithmic}
\end{algorithm}

\begin{algorithm}[ht]
\caption{Joint state-parameter estimation by Particle MCMC}\label{alg:optSPF_MCMC}
\begin{algorithmic}[1]
\REQUIRE $y_{1:T}$, $M$, $\btheta^{(0)}$, $\bm{\Sigma}^{(0)}$,
$\gamma\in(1/2,1]$, $\bar{\alpha}_\ast$, and $\varphi_T(\cdot)$
\ENSURE $\hat{\bfx}_{1:T}$ and $\hat{\btheta}$ \STATE Initialize:
$\mathbf{S}_0 = \mathrm{Chol}\left( \bm{\Sigma}^{(0)} \right)$ and
$\hat{\varphi}_T(\btheta^{(0)})=0$ \FOR{$j=1$ to $M$} \STATE Draw
$\mathbf{a} \sim \mathcal{N} (\mathbf{0}, \mathbf{I} )$ \STATE
Compute $\btheta^{\ast} = \btheta^{(j-1)} + \mathbf{S}_{j-1}
\mathbf{a} $
\STATE Run the PF in Algorithm \ref{alg:optSPF} with model parameters set to $\btheta^{\ast}$. \\
Required outputs: \\
~~~~ State filtering $\hat{\bfx}_{1:T}^\ast$ as in (\ref{eq:MMSEpf2}) \\
~~~~ Energy function $\hat{\varphi}_T(\btheta^{\ast})$ as in (\ref{eq:enfun_est})
\STATE Compute $\alpha^{(j)} = \min \left\{ 1, \exp(\hat{\varphi}_T(\btheta^{(j-1)}) - \hat{\varphi}_T(\btheta^{\ast})) \right\}$
\STATE Draw $u \sim \mathcal{U}(0,1)$
\IF{$u\leq \alpha^{(j)}$}
\STATE $\btheta^{(j)} = \btheta^\ast$
\STATE $\hat{\bfx}_{1:T} = \hat{\bfx}_{1:T}^\ast$
\ELSE
\STATE  $\btheta^{(j)} = \btheta^{(j-1)}$
\ENDIF
\STATE Compute $\eta^{(j)} = j^{-\gamma}$
\STATE Compute $\mathbf{D}_{j} = \left( \mathbf{I} + \eta^{(j)} (\alpha^{(j)} - \bar{\alpha}_\ast) \frac{\mathbf{a}\mathbf{a}^\top}{||\mathbf{a}||^2}\right)$
\STATE Compute $\mathbf{S}_{j} = \mathrm{Chol} \left( \mathbf{S}_{j-1} \mathbf{D}_{j} \mathbf{S}_{j-1}^\top \right)$
\ENDFOR
\STATE State filtering $\Rightarrow \hat{\bfx}_{1:T}$
\STATE Parameter estimation with $\left\{\btheta^{(j)}\right\}_{j=1}^{M}$ as in (\ref{eq:theta_est}) $\Rightarrow \hat{\btheta}$
\end{algorithmic}
\end{algorithm}

\section{Computer simulation results} \label{sect:seq_estimation:results}

We simulated data of a neuron following the Morris-Lecar model.
Particularly, we generated data sampled at $f_s=4$ kHz. The model
parameters were set to $C_m=20$ $\mu$F/cm$^2$, $\phi=0.04$,
$V_1=-1.2$ mV, $V_2=18$ mV, $V_3=2$ mV, and $V_4=30$ mV; the reverse
potentials were $E_\mathrm{L}=-60$ mV, $E_\mathrm{Ca}=120$ mV, and
$E_\mathrm{K}=-84$ mV; and the maximal conductances were
$\bar{g}_\mathrm{Ca}=4.4$ mS/cm$^2$ and $\bar{g}_\mathrm{K}=8.0$
mS/cm$^2$. We considered a measurement noise with a standard
deviation of $1$ mV, which corresponds to an SNR of $32$ dB. This
value is considered a reasonable value in nowadays intracellular
sensing devices. Model inaccuracies were generated as in Section
\ref{sect:seq_estimation:inaccuracies}.

Three sets of simulations are discussed. First, we validated the
filtering method considering perfect knowledge of the model. In this
case, the method in Algorithm \ref{alg:optSPF} was used. Secondly,
the model assumptions were relaxed in the sense that the parameters
of the model were not known by the method. We analyzed the
capabilities of the proposed method to infer both the time-evolving
states of the system and some of the parameters defining the model.
In this case, the method in Algorithm \ref{alg:optSPF_MCMC} was
used. Finally, we validated the performance of the proposed methods
in inferring the synaptic conductances. We tested both \ac{PF} and
\ac{PMCMC} methods, that is with and without full knowledge of the
model respectively.

\subsection{Model parameters are known}
\label{sect:seq_estimation:results:opt}

In the simulations we considered the model inaccuracies described in
Section \ref{sect:seq_estimation:inaccuracies}. To excite the neuron
into spiking activity a nominal applied current was injected with
$I_o=110$ $\mu$A/cm$^2$ and two values for $\sigma_I$ were
considered, namely $1\%$ and $10\%$ of $I_o$. The nominal
conductance used in the model was $\bar{g}_\mathrm{L}=2$ mS/cm$^2$,
whereas the underlying neuron had a zero-mean Gaussian error with
standard deviation $\sigma_{\bar{g}_\mathrm{L}}$. Two variance
values were considered as well, $1\%$ and $10\%$ of
$\bar{g}_\mathrm{L}$. Finally, we considered $\sigma_n=10^{-3}$ in
the dynamics of the gating variable.

In order to evaluate the efficiency of the proposed estimation
method, we computed the \ac{PCRB} \cite{VanTrees07} in
Appendix \ref{ap:seq_estimation:bounds}. We plot the \ac{PCRB} as a
benchmark for the \ac{RMSE} curves obtained by computer simulations, obtained
after averaging $200$ independent Monte Carlo trials. For a generic time
series $w_k$, the \ac{RMSE} of an estimator $\hat{w}_k$ is defined
as
\begin{equation}
\nonumber  \textrm{RMSE}(w_k) =\sqrt{\mathbb{E}\{(w_k - \hat{w}_k )^2\}}
  \approx \sqrt{\frac{1}{M} \sum_{j=1}^{M} (w_k - \hat{w}_{j,k} )^2
  } ~,
\end{equation}
\noindent where $\hat{w}_{j,k}$ denotes the estimate of $w_k$ at the
$j$-th realization and $M$ the number of independent Monte Carlo
trials used to approximate the mathematical expectation.

\begin{figure}[ht]
 \begin{center}
  \includegraphics[width=8cm]{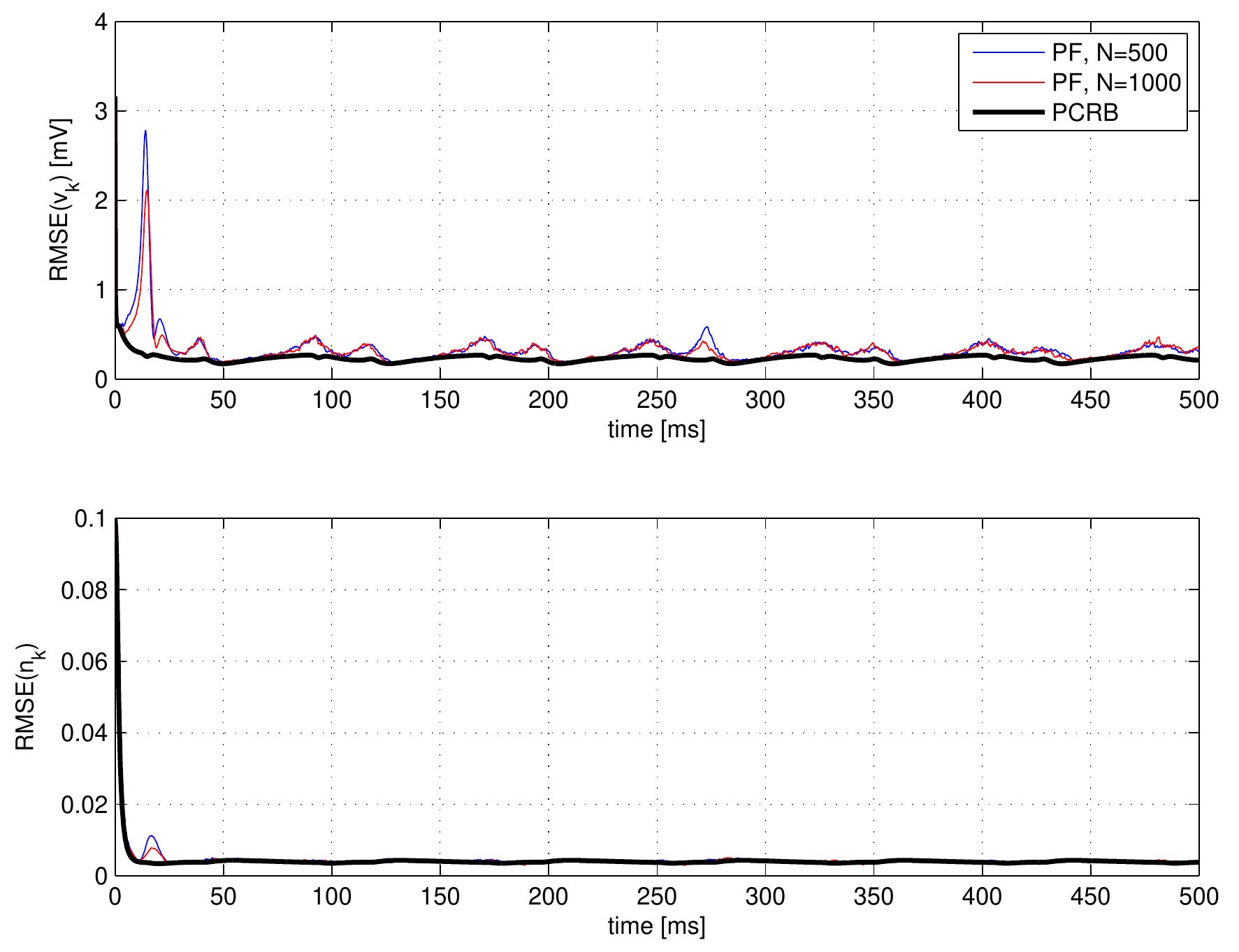}
  \caption{Evolution of the RMSE and the PCRB over time. Model inaccuracies where
  $\sigma_I = 0.01 \cdot I_o$ and $\sigma_{g_\mathrm{L}} = 0.01 \cdot \bar{g}_\mathrm{L}^o$.}\label{fig:RMSE_SNR32_1per100}
 \end{center}
\end{figure}

\begin{figure}[ht]
 \begin{center}
  \includegraphics[width=8cm]{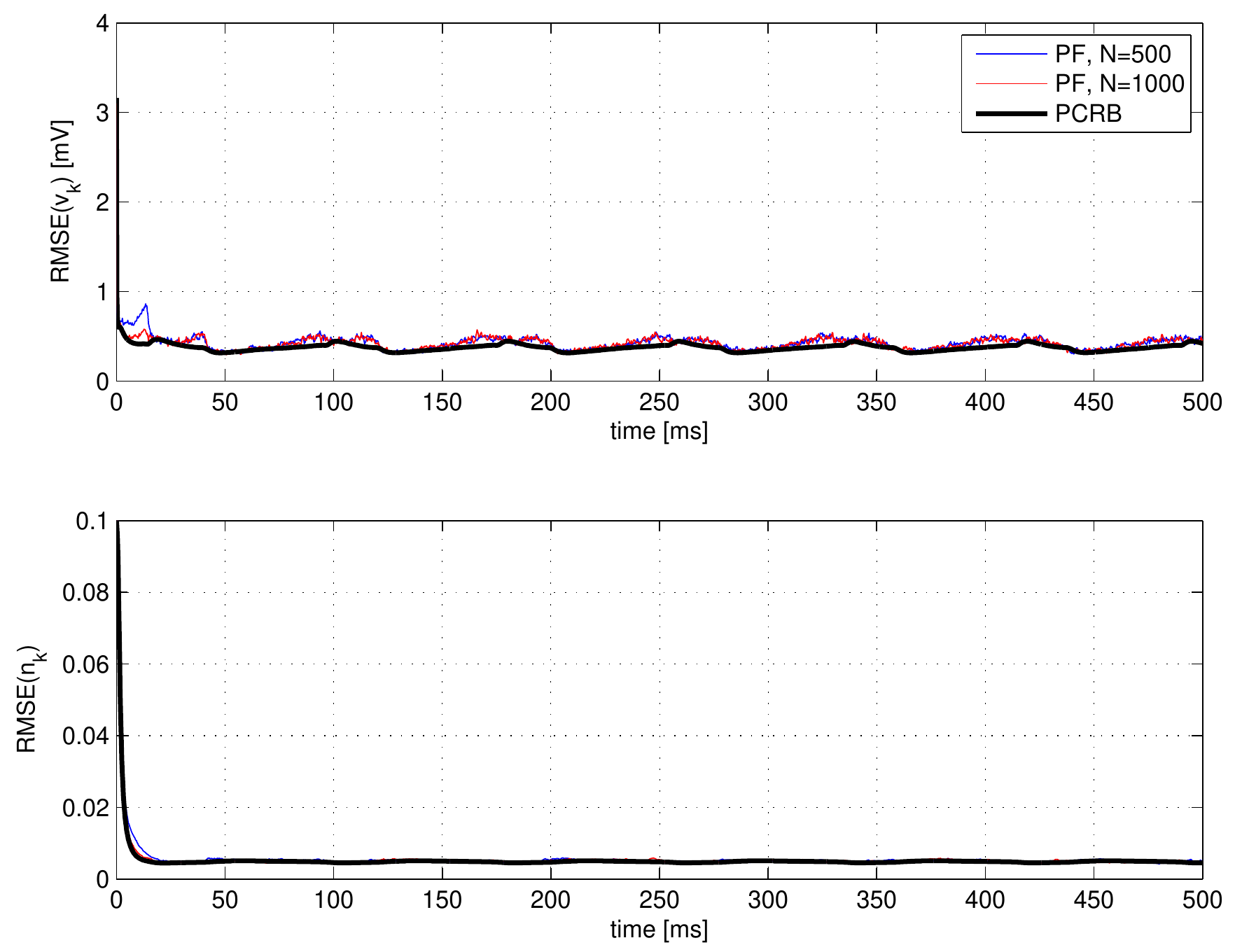}
  \caption{Evolution of the RMSE and the PCRB over time. Model inaccuracies where
  $\sigma_I = 0.1 \cdot I_o$ and $\sigma_{g_\mathrm{L}} = 0.1 \cdot \bar{g}_\mathrm{L}^o$.}\label{fig:RMSE_SNR32_10per100}
 \end{center}
\end{figure}

Figures \ref{fig:RMSE_SNR32_1per100} and
\ref{fig:RMSE_SNR32_10per100} show the time course of the \ac{RMSE}
using $N=\{500,1000\}$ particles. We see that in both scenarios, our
method efficiently attains the \ac{PCRB}. We measure the efficiency
($\eta \geq 1$) of the method as the quotient between the \ac{RMSE}
and the \ac{PCRB}, averaged over the entire simulation time. The
worse efficiency on estimating $v_k$ was $1.43$ corresponding to 500
particles and $10\%$ of inaccuracies (see Fig.
\ref{fig:RMSE_SNR32_10per100}), the best was $1.11$ for $1000$
particles and $1\%$ of errors (see Fig.
\ref{fig:RMSE_SNR32_1per100}). In estimating $n_k$ the discrepancy
was even lower, $1.06$ and $1.03$ for maximum and minimum $\eta$. As
a conclusion, the \ac{PF} tends to the \ac{PCRB} with the number of
particles. Also, the performance (both theoretical and empirical)
could be improved if model inaccuracies are reduced, i.e., if the
model parameters are better estimated at a previous stage. For the
sake of completeness, we summarize the results in Table
\ref{table:results_optPF}, where the average \ac{RMSE} and \ac{PCRB}
along the $500$ ms simulation can be consulted. It is apparent that
increasing the number of particles from $N=500$ to $N=1000$ does not
improve significantly the performance of the method.

\begin{table}[ht]
\begin{center}
\begin{tabular}{|c|c|c|c|c|}
  \hline
  \multirow{2}{*}{}& \multicolumn{2}{c|}{$\sigma_I = 0.01 \cdot I_o$,} & \multicolumn{2}{c|}{$\sigma_I = 0.1 \cdot I_o$,} \\
  \multirow{2}{*}{}& \multicolumn{2}{c|}{$\sigma_{g_\mathrm{L}} = 0.01 \cdot \bar{g}_\mathrm{L}^o$} & \multicolumn{2}{c|}{$\sigma_{g_\mathrm{L}} = 0.1 \cdot \bar{g}_\mathrm{L}^o$} \\
  \cline{2-5}
                   & $N=500$ & $N=1000$ & $N=500$ & $N=1000$ \\ \hline
  $\langle$RMSE($v_k$)$\rangle$ & 0.3344 & 0.3211 & 0.4269 & 0.4203 \\
  $\langle$PCRB($v_k$)$\rangle$ & 0.2325 & 0.2325 & 0.3777 & 0.3777 \\
  $\langle$RMSE($n_k$)$\rangle$ & 0.0046 & 0.0045 & 0.0056 & 0.0055 \\
  $\langle$PCRB($n_k$)$\rangle$ & 0.0043 & 0.0043 & 0.0053 & 0.0053 \\
  \hline
\end{tabular}
\caption{Averaged results over simulation time.}\label{table:results_optPF}
\end{center}
\end{table}

\begin{figure}[ht]
 \begin{center}
     \subfigure[$\mathrm{SNR}=0$ dB]{\scalebox{0.75}{\includegraphics[width=10cm]{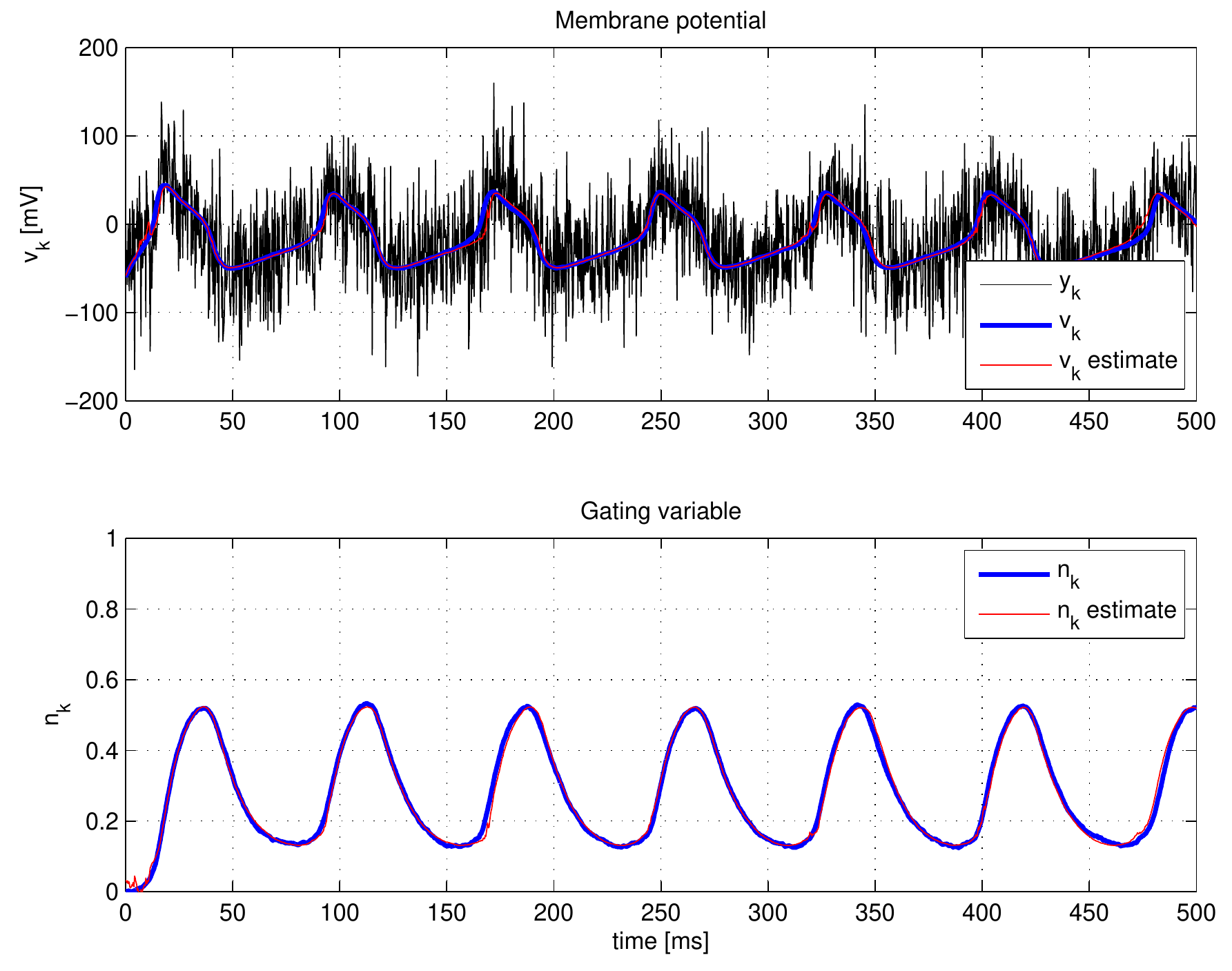}}\label{fig:ML_SNR0_N500}} \\
     \subfigure[$\mathrm{SNR}=32$ dB]{\scalebox{0.75}{\includegraphics[width=10cm]{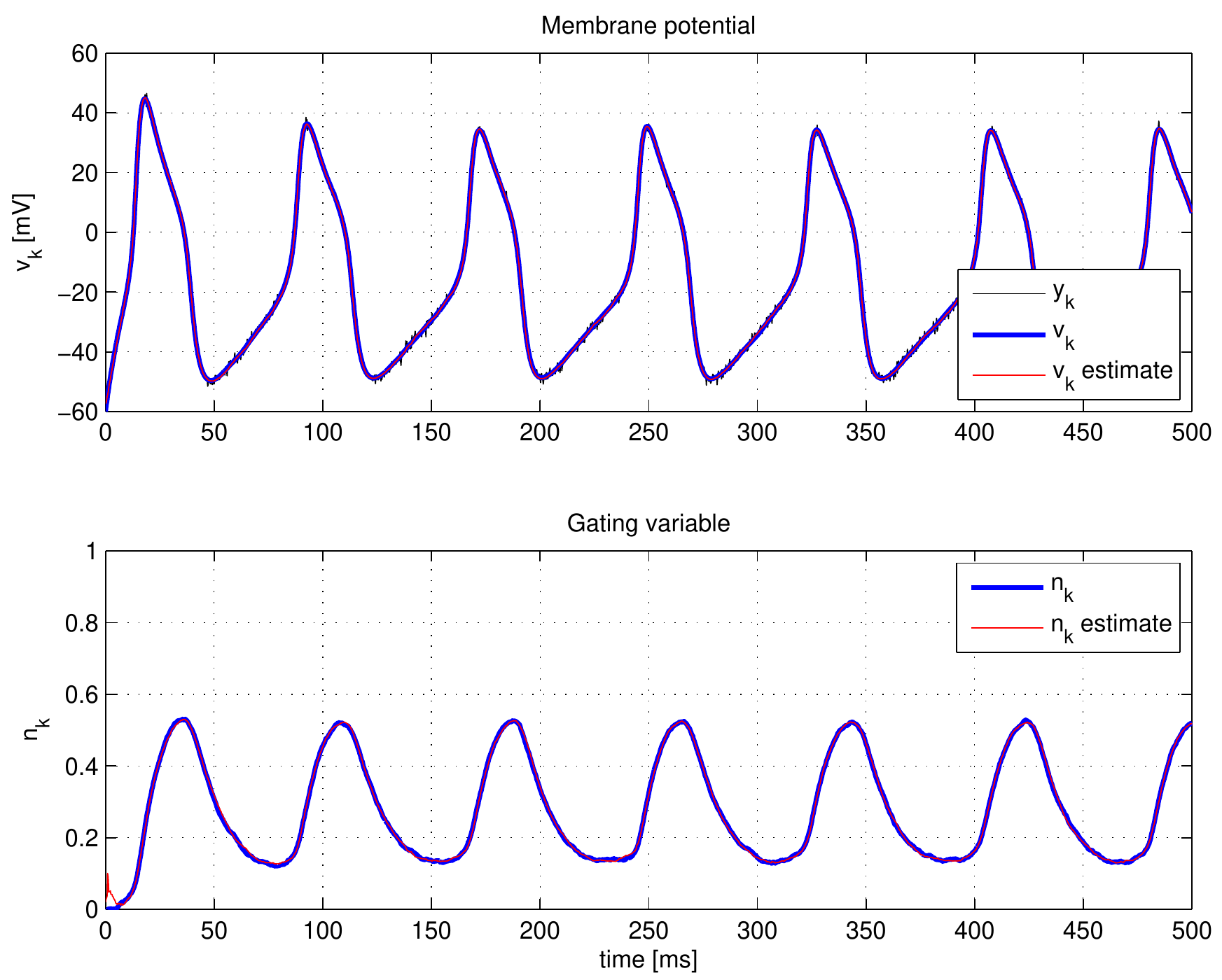}}\label{fig:ML_SNR32_N500}}
    \caption{A single realization of the \ac{PF} method for (a) $\mathrm{SNR}=0$ dB and (b) $\mathrm{SNR}=32$ dB.}\label{fig:ML_N500}
 \end{center}
\end{figure}

To give some intuition on the operation and performance of the
\ac{PF} method in Algorithm \ref{alg:optSPF}, we show the results
for a single realization in Fig. \ref{fig:ML_N500}. The results are for 500 particles and two
different values of $\sigma_{y,k}^2$, corresponding to $0$ and $32$
dB respectively. Even in very low SNR regimes, the method is able to
operate and provide reliable filtering results.

\subsection{Model parameters are unknown}
\label{sect:seq_estimation:results:PFMCMC}

In this section we validate the algorithm presented in Section
\ref{subsect:seq_estimation:method2}. According to the previous
analysis, we deem that $500$ particles are enough for the filter to
provide reliable results. The parameters of the \ac{PMCMC} algorithm
were set to $\gamma=0.9$ and $\bar{\alpha}_\ast = 0.234$.

\begin{figure*}[ht]
 \begin{center}
     \subfigure[$\bar{g}_\mathrm{Ca}$]{\scalebox{1}{\includegraphics[width=7cm]{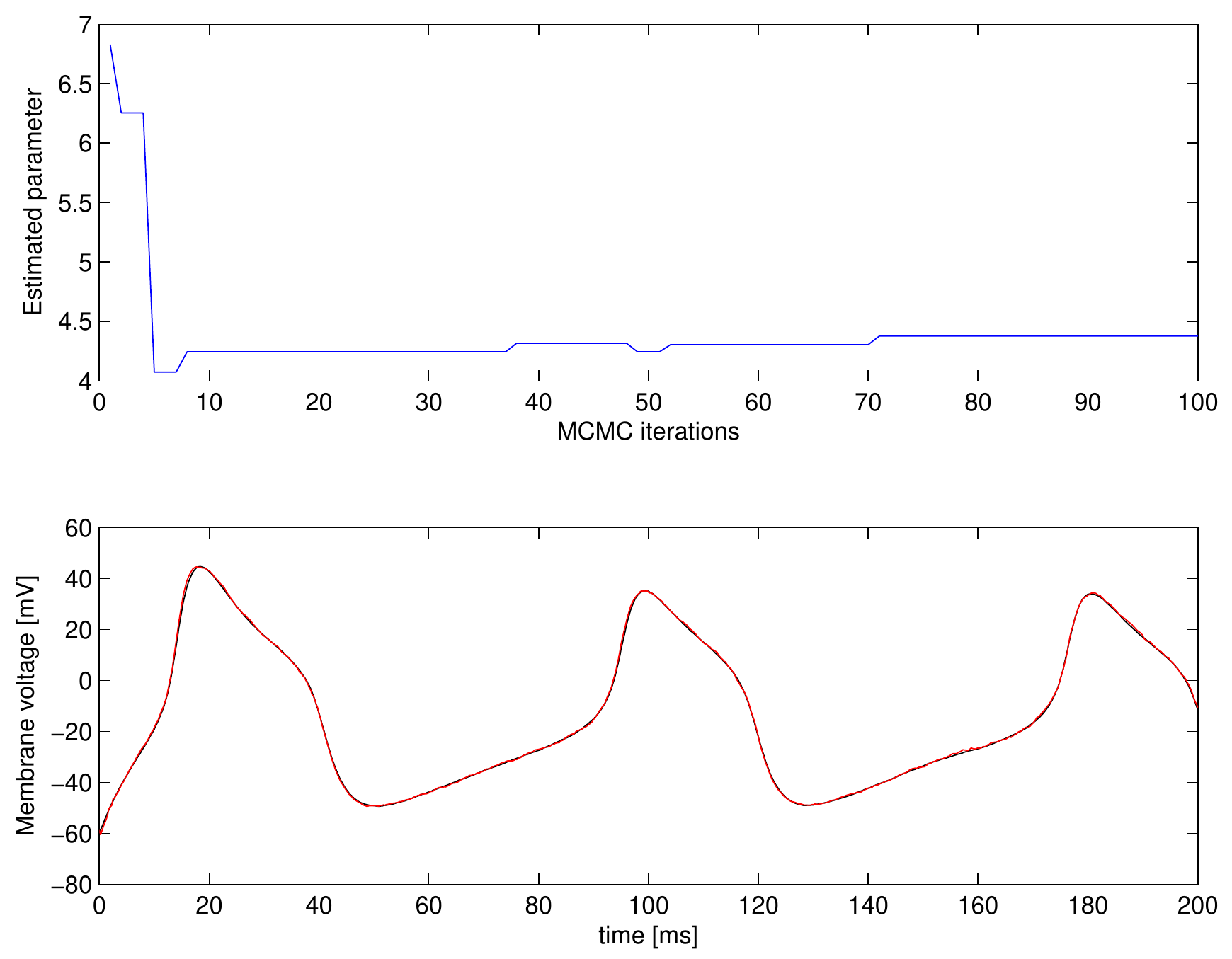}}\label{fig:PMCMC_gca}}
     \subfigure[$\bar{g}_\mathrm{K}$]{\scalebox{1}{\includegraphics[width=7cm]{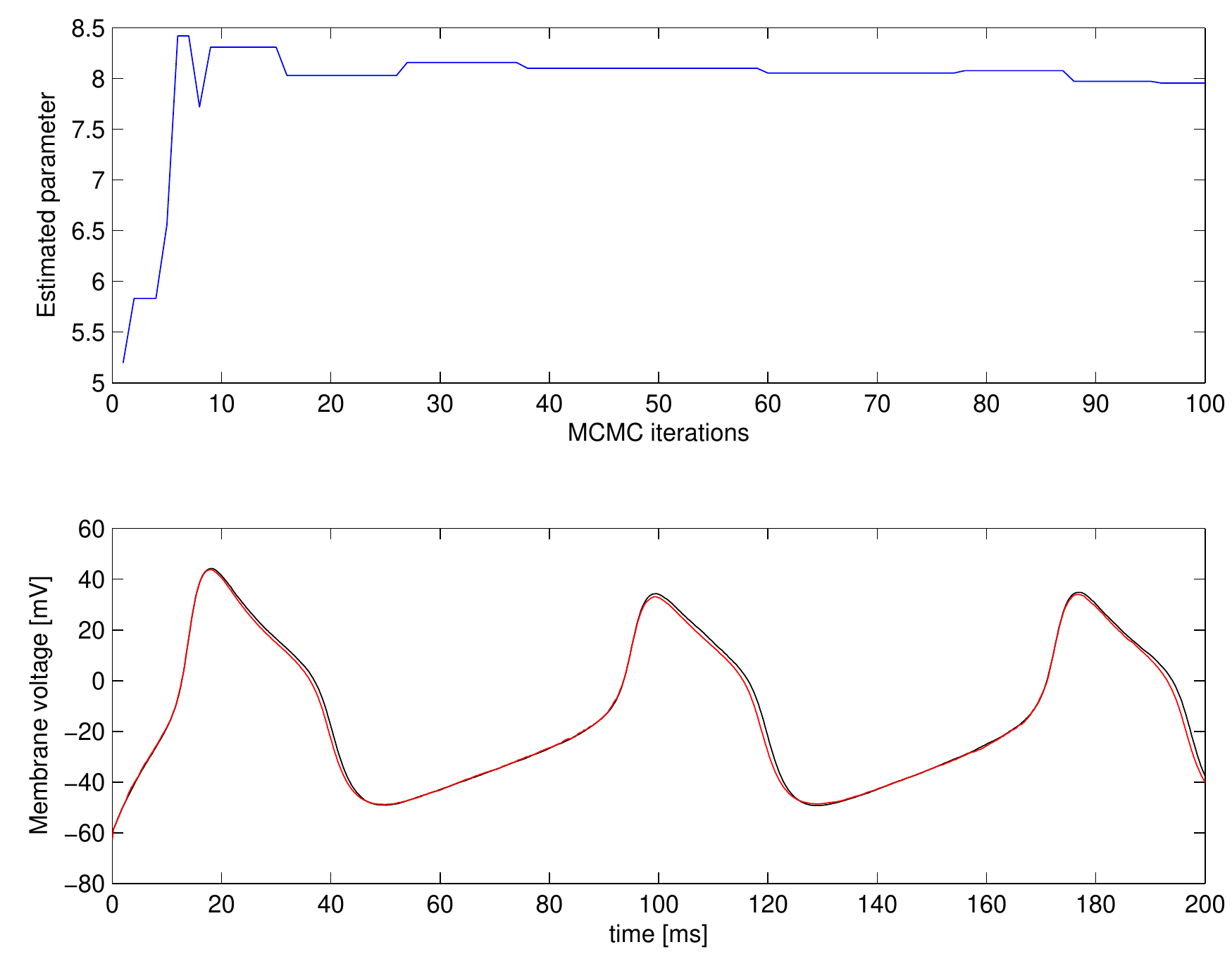}}\label{fig:PMCMC_gk}} \\
     \subfigure[$\bar{g}_\mathrm{Ca}$ and $\bar{g}_\mathrm{K}$]{\scalebox{1}{\includegraphics[width=7cm]{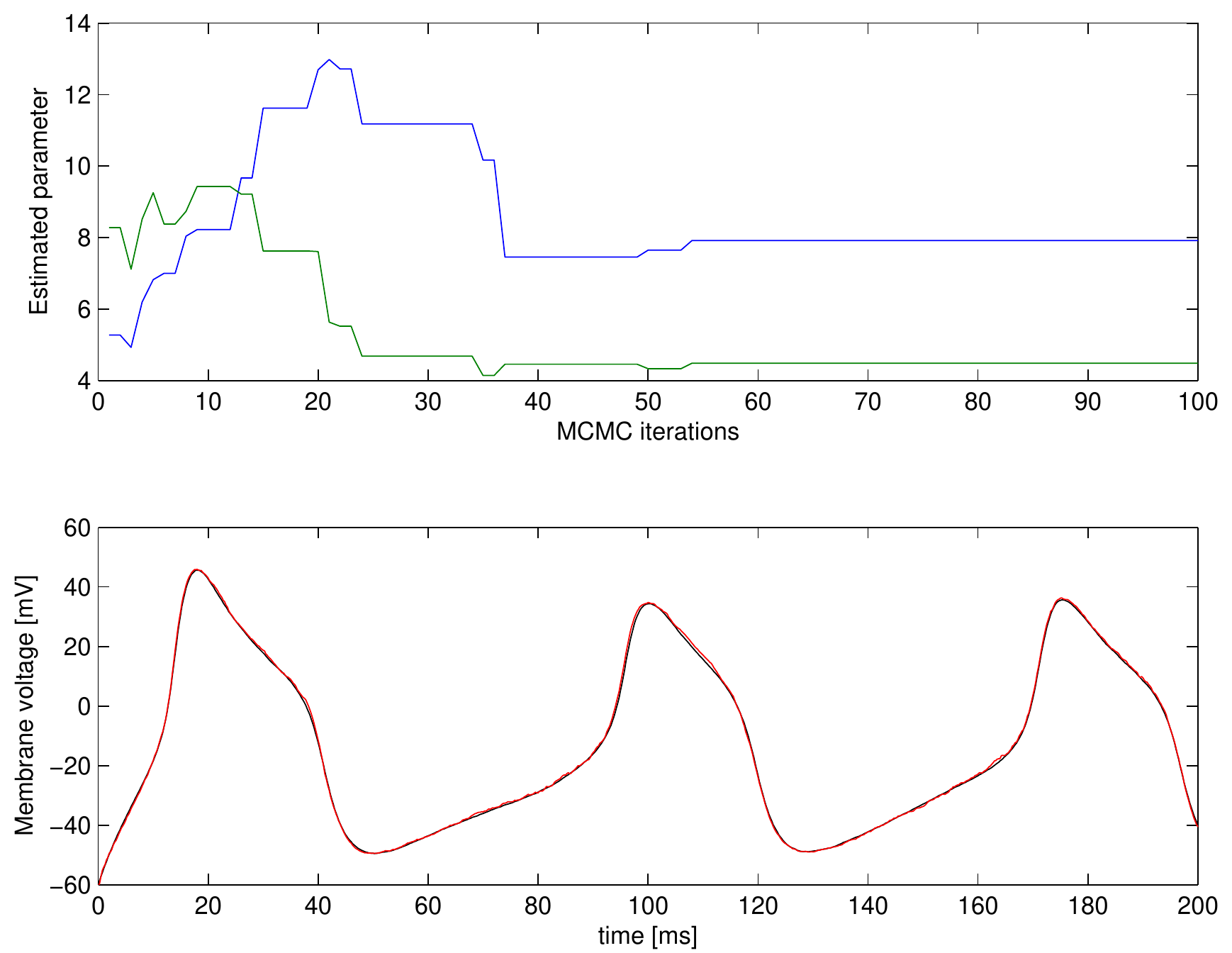}}\label{fig:PMCMC_gk_gca}}
     \subfigure[$\bm{\Sigma}_{x,k}^{1/2}=\textrm{diag}(\sigma_v,\sigma_n)$]{\scalebox{1}{\includegraphics[width=7cm]{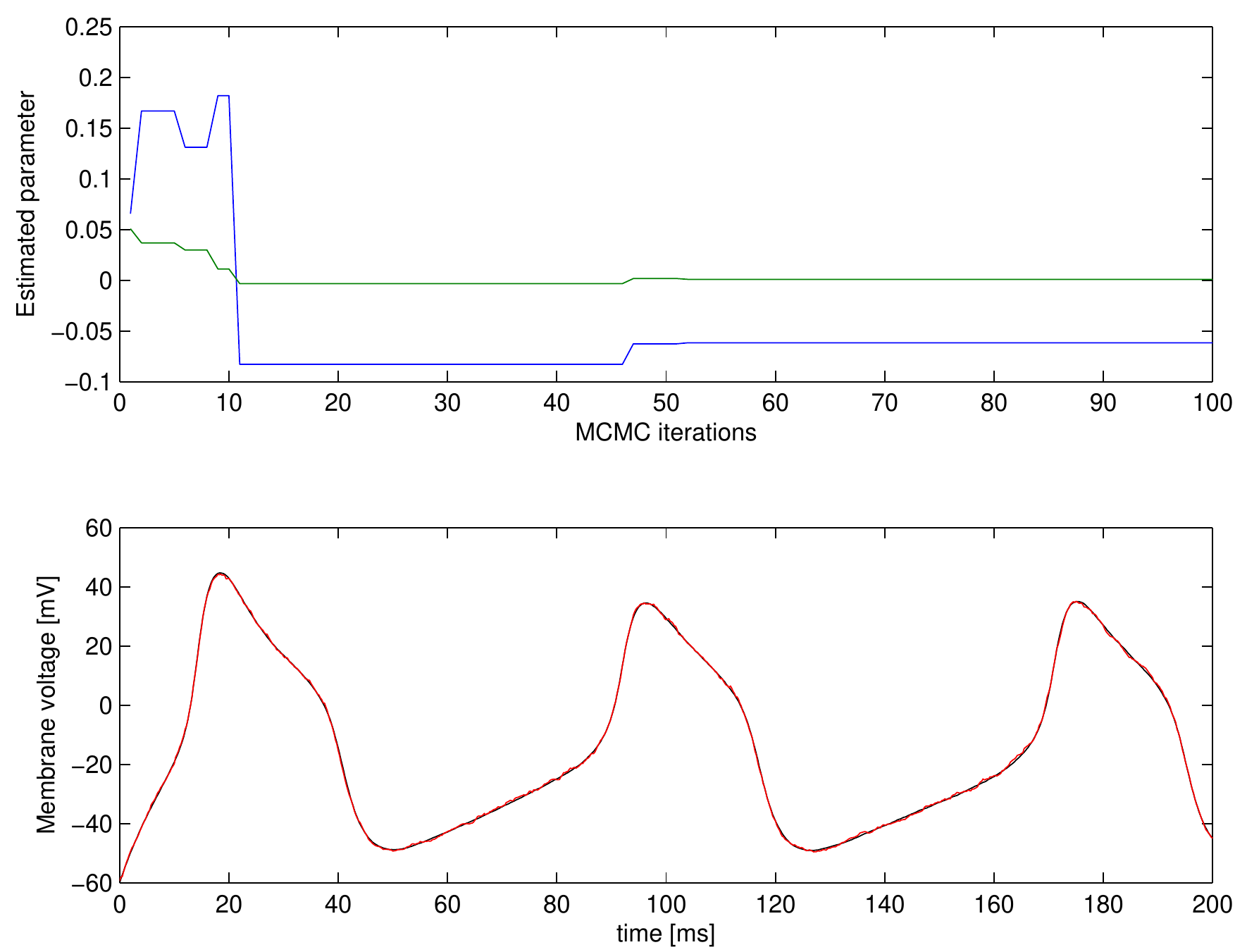}}\label{fig:PMCMC_sigmax}} \\
     \subfigure[$\sigma_{y,k}$]{\scalebox{1}{\includegraphics[width=7cm]{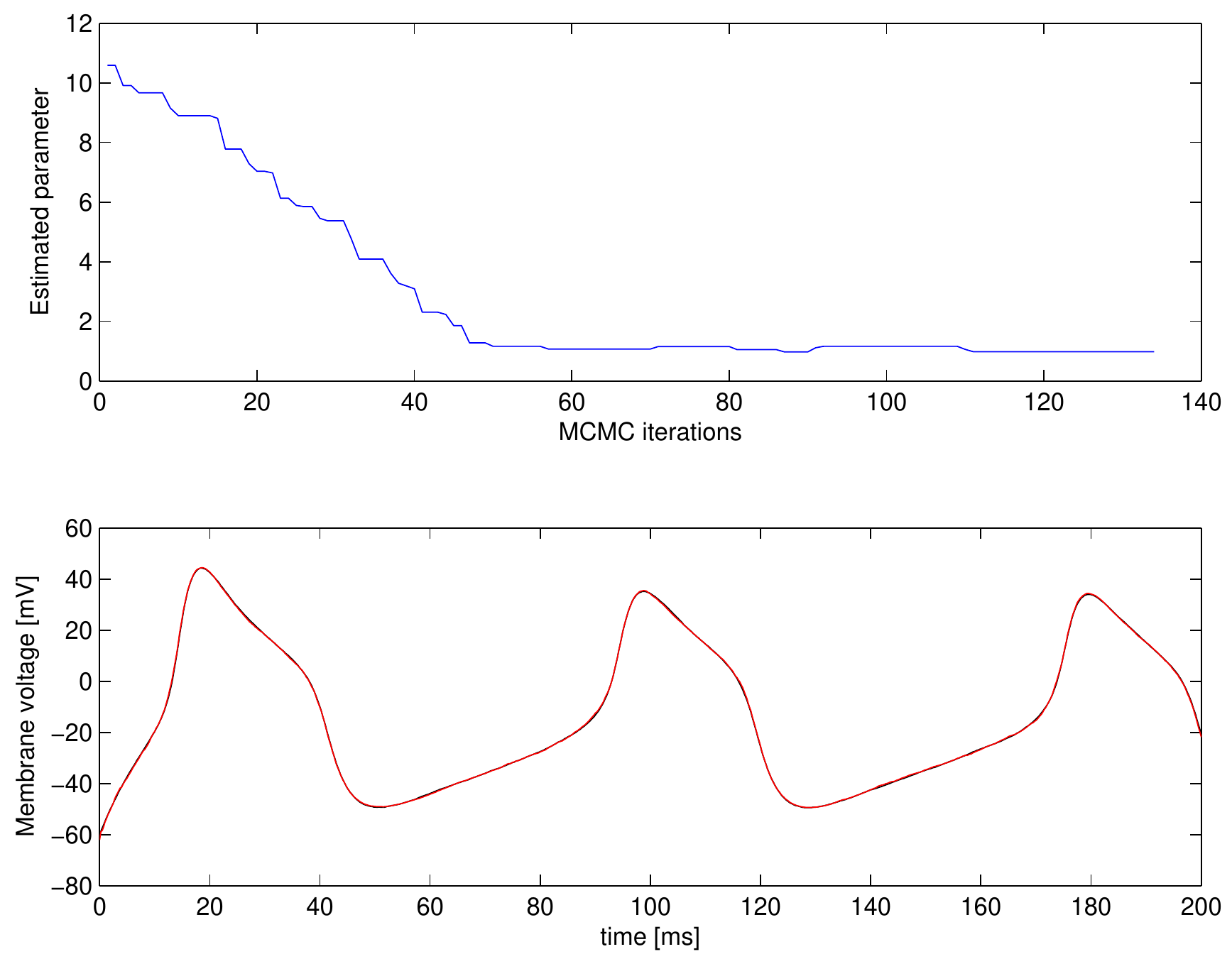}}\label{fig:PMCMC_sigmay}}
     \subfigure[$\bar{g}_\mathrm{Ca}$, $\bar{g}_\mathrm{K}$, and $\bm{\Sigma}_{x,k}^{1/2}$]{\scalebox{1}{\includegraphics[width=7cm]{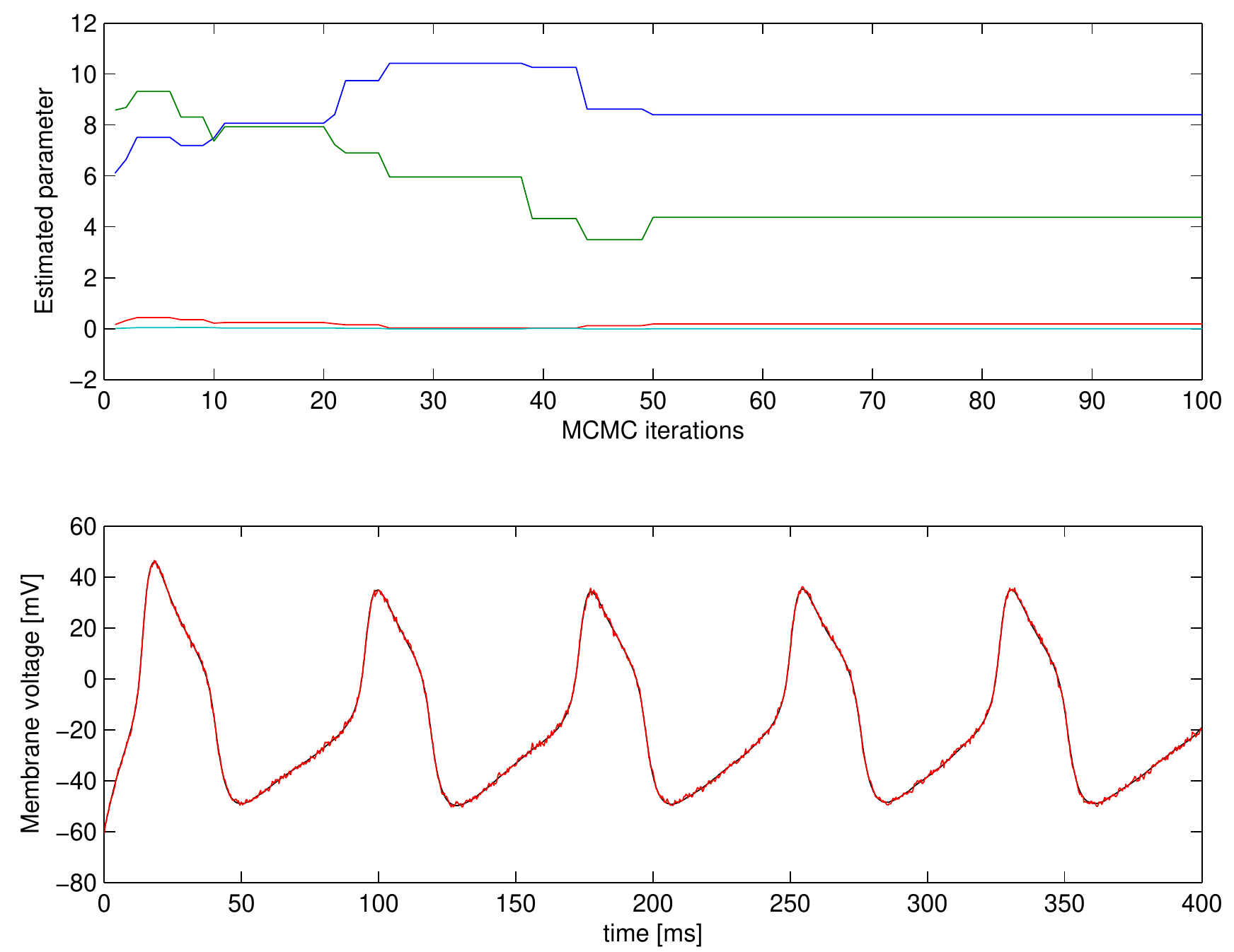}}\label{fig:PMCMC_gk_gca_sigmax}} \\
    \caption{Realizations of the PMCMC algorithm for joint state-parameter estimation. Each plot corresponds to different unknown parameters, featuring the MCMC iterations (top) that converge to the true value of the parameter and the filtered voltage trace (bottom).}\label{fig:PMCMC_res}
 \end{center}
\end{figure*}

Fig. \ref{fig:PMCMC_res} shows the results for a single realization
when a number of parameters in the nominal model are unknown. We
considered 1, 2, and 4 unknown parameters. Each of the plots include
$M=100$ iterations of the \ac{MCMC} showing the evolution of the
parameter estimation (top) and the superimposed recorded voltage in
black and the filtered voltage trace in red (bottom). Model
inaccuracies are of $1\%$, similarly as in Fig.
\ref{fig:RMSE_SNR32_1per100}. In these plots we omitted the
results for the gating variable for the sake of clarity. The true
and initial values used in the experiments, as well as the initial
covariances assumed, can be consulted in Table
\ref{table:PMCMC_res_values}. From the plots we can observe that the
method performs reasonably well even in the case of estimating the
model parameters at the same time it is filtering out the noise in
the membrane voltage traces.

\begin{figure}[ht]
 \begin{center}
     \subfigure[]{\scalebox{0.75}{\includegraphics[width=7.5cm]{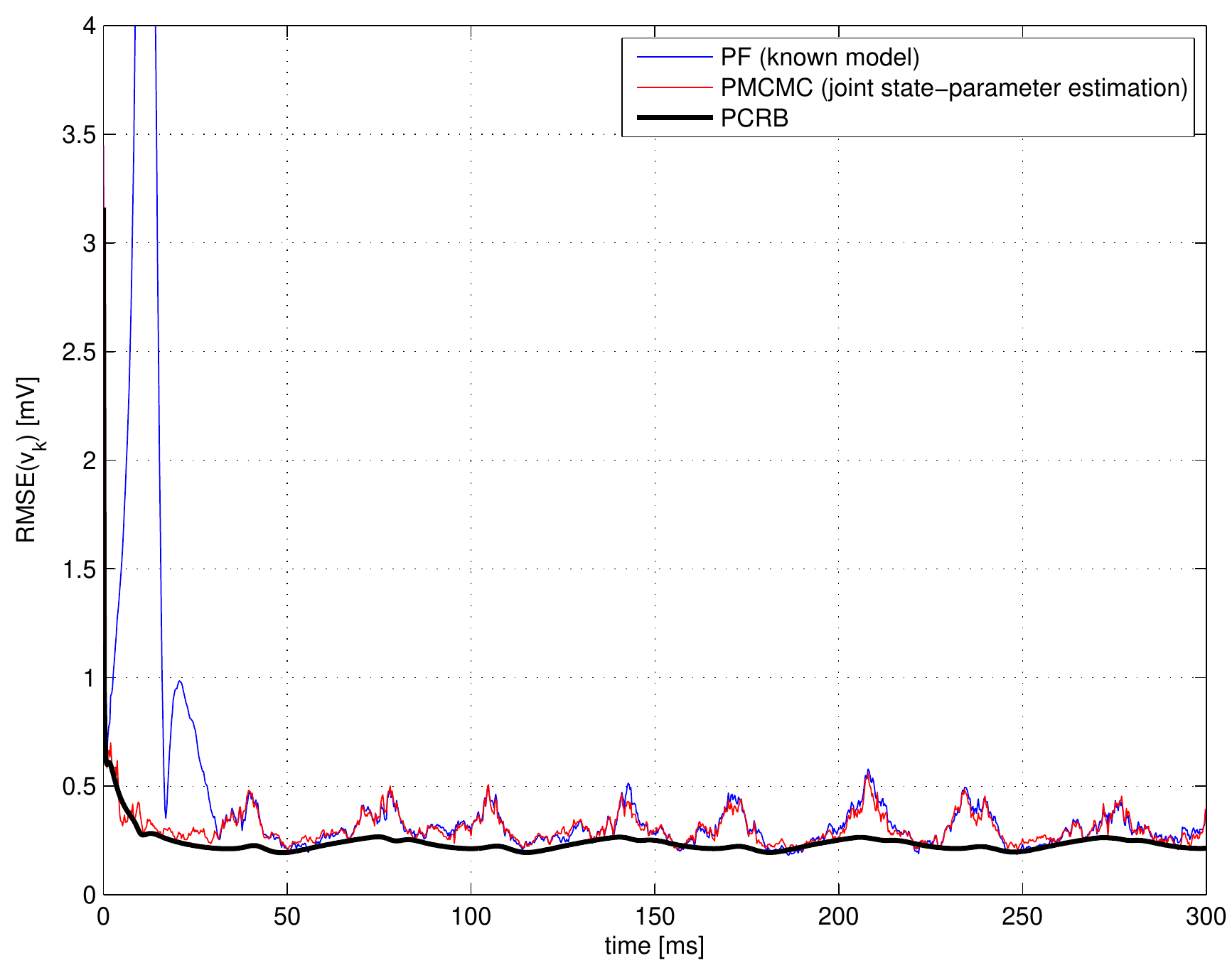}}\label{fig:RMSEv_N500_PMCMC_1per100}} \\
     \subfigure[]{\scalebox{0.75}{\includegraphics[width=7.5cm]{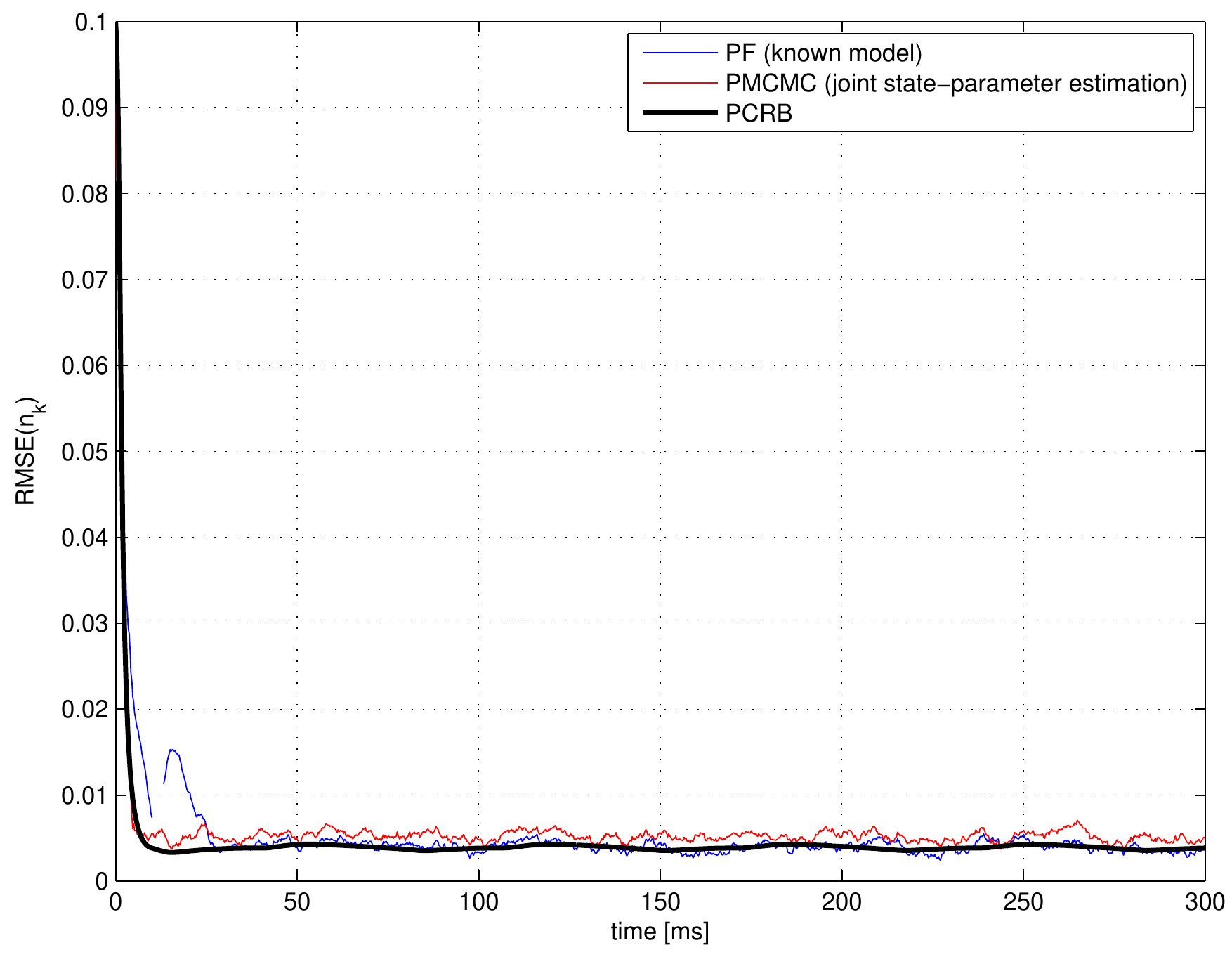}}\label{fig:RMSEn_N500_PMCMC_1per100}}
  \caption{Evolution of RMSE($v_k$) and RMSE($n_k$) over time for the PMCMC method estimating the leakage parameters. Model inaccuracies where
  $\sigma_I = 0.1 \cdot I_o$ and $\sigma_{g_\mathrm{L}} = 0.1 \cdot \bar{g}_\mathrm{L}^o$.}\label{fig:RMSE_N500_PMCMC_1per100}
 \end{center}
\end{figure}

A biologically meaningful signal is the leakage current. In
general, the leakage gathers those ionic channels that are not
explicitly modeled and other non-modeled sources of activity. The
parameters driving the leak current are $\bar{g}_L$ and
$E_\mathrm{L}$. We tested and validated the proposed
\ac{PMCMC} in an experiment where the leak parameters were estimated
at the same time the filtering solution was computed. Moreover, the
statistics of the process noise were estimated as well,
$\bm{\Sigma}_{x,k}$. In this case, we iterated the \ac{PMCMC} method
1000 times and average the results over 100 Monte Carlo independent
trials. The results can be consulted in Fig.
\ref{fig:RMSE_N500_PMCMC_1per100}, where the \ac{RMSE} performance
of the \ac{PMCMC} method is compared to the performance of the
original \ac{PF} with perfect knowledge of the model.

It can be observed that the filtering performances with perfect
knowledge of the model and with estimation of parameters by
\ac{PMCMC} are similar. Moreover, both approaches attain the
theoretical lower bound of accuracy given by the \ac{PCRB}.

\begin{figure}[ht]
 \begin{center}
     \subfigure[$\bar{g}_\mathrm{L}$ estimates]{\scalebox{0.75}{\includegraphics[width=7cm]{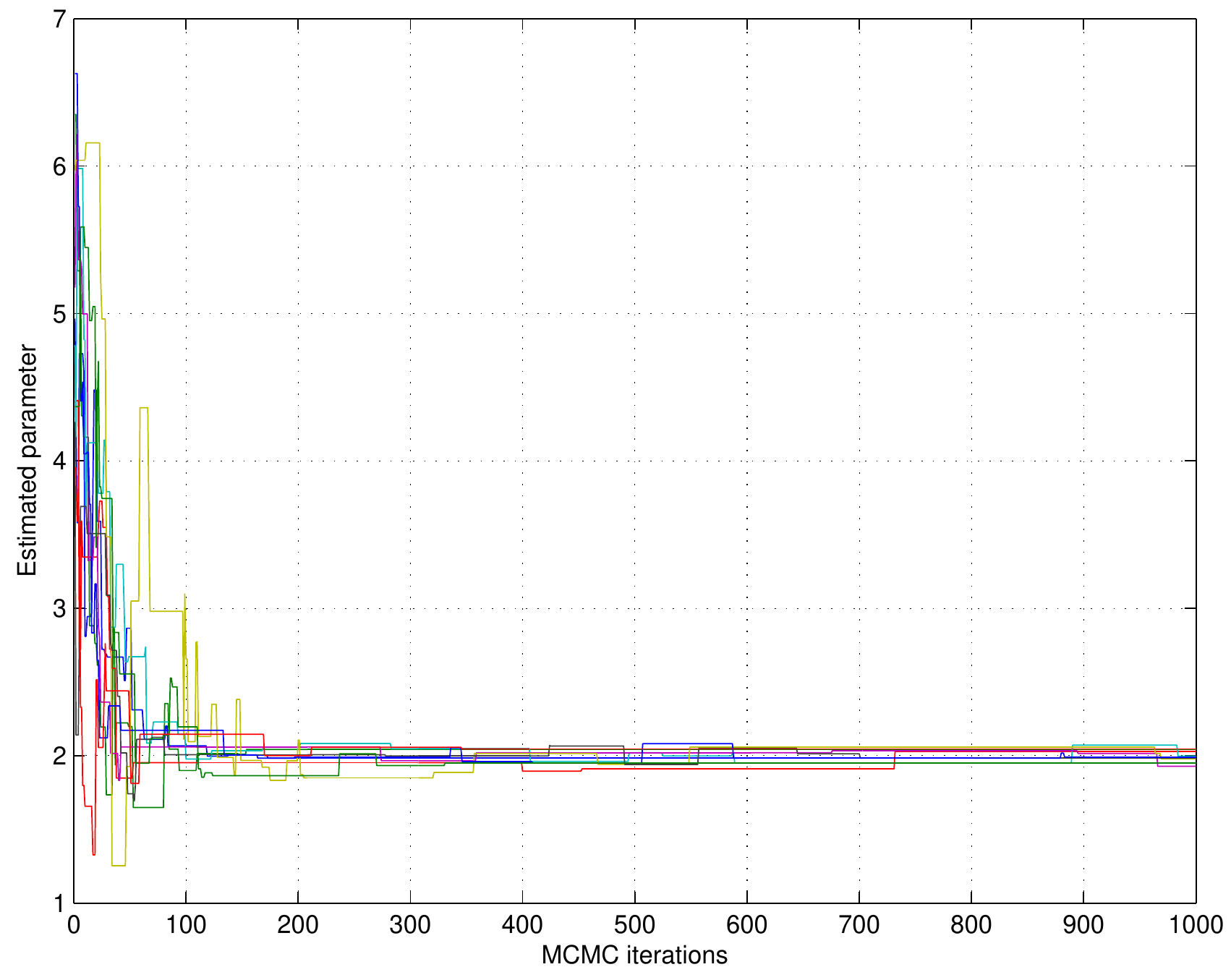}}\label{fig:PMCMC_res_param1}}
     \subfigure[Mean $\bar{g}_\mathrm{L}$ estimate]{\scalebox{0.75}{\includegraphics[width=7cm]{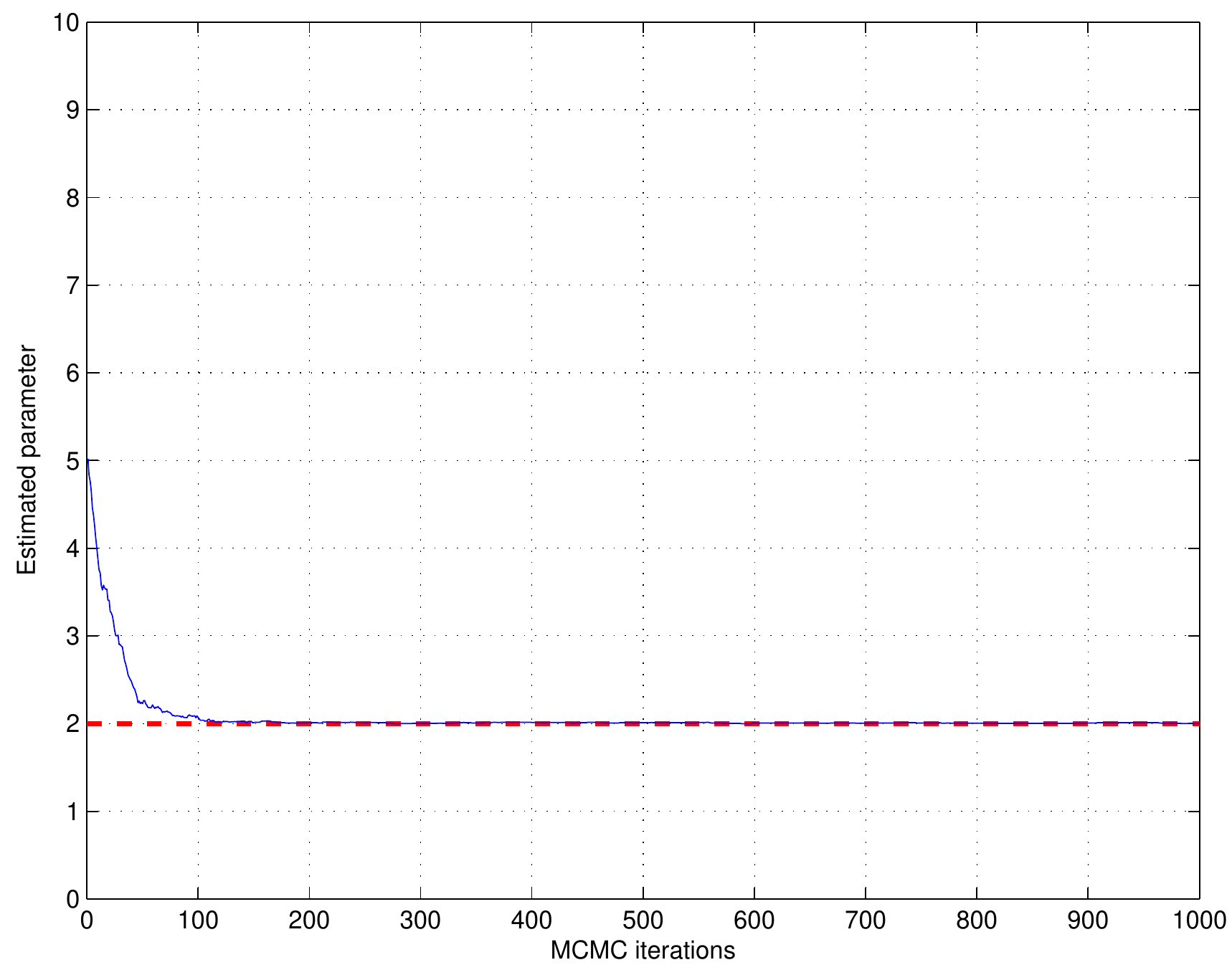}}\label{fig:PMCMC_res_param2}} \\
     \subfigure[$E_\mathrm{L}$ estimates]{\scalebox{0.75}{\includegraphics[width=7cm]{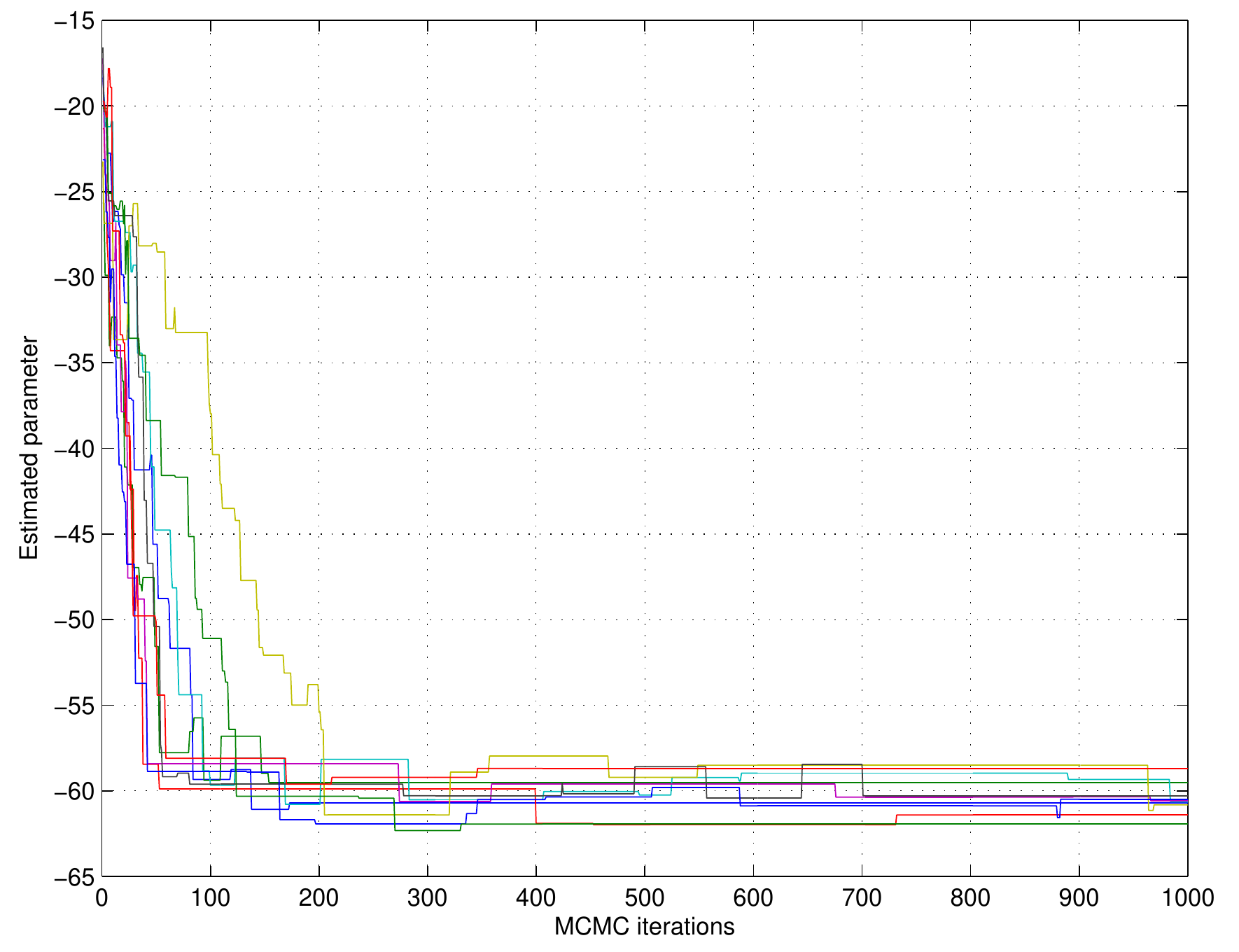}}\label{fig:PMCMC_res_param3}}
     \subfigure[Mean $E_\mathrm{L}$ estimate]{\scalebox{0.75}{\includegraphics[width=7cm]{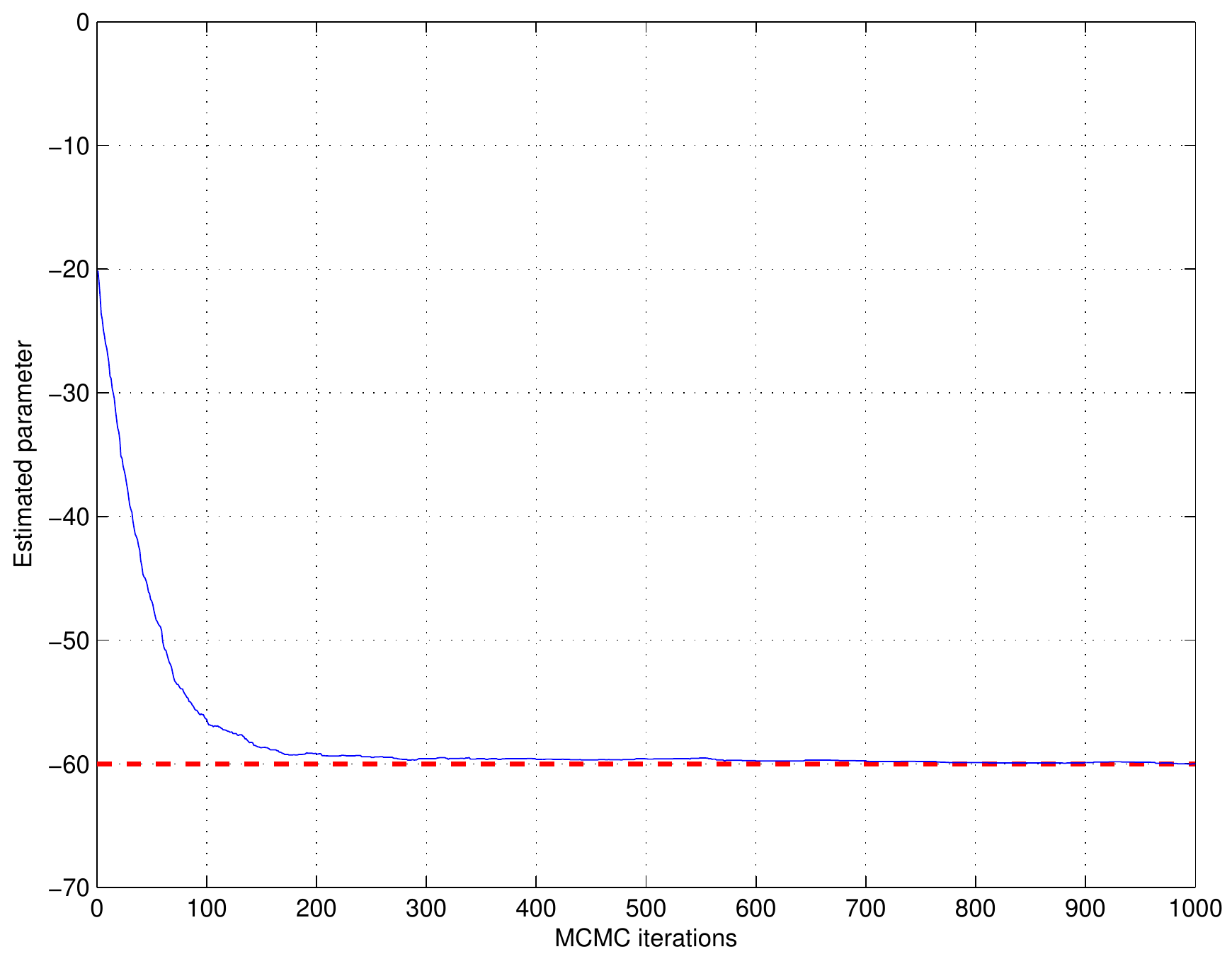}}\label{fig:PMCMC_res_param4}} \\
    \caption{Parameter estimation performance of the proposed PMCMC algorithm. Top plots show results for $\bar{g}_\mathrm{L}=2$ estimation and bottom plots for $E_\mathrm{L}=-60$. Plots (b) and (d) show superimposed independent realizations and plots (a) and (c) show the average estimate of the parameter.}\label{fig:PMCMC_res_param}
 \end{center}
\end{figure}

In Fig. \ref{fig:PMCMC_res_param}, validation results for the
parameter estimation capabilities of the \ac{PMCMC} are shown.
Particularly, we plotted in Fig. \ref{fig:PMCMC_res_param1} and
\ref{fig:PMCMC_res_param3} a number of independent realizations of
the samples trajectories  $\left\{\btheta^{(j)}\right\}_{j=1}^{M} $.
We observe that all of them converge to the true values of the
parameter. Recall that these true values were $\btheta = (\bar{g}_L
, E_\mathrm{L})^\top = (2,-60)^\top$. In Fig.
\ref{fig:PMCMC_res_param2} and \ref{fig:PMCMC_res_param4}, the
average of these realizations can be consulted, where the
aforementioned convergence to the true parameter is highlighted.

\begin{table}[t]
\centering
\begin{tabular}{|c|c|c|c|c|}
  \hline
    Parameter & True value & Initial value & Init. Covariance \\ \hline
    $\bar{g}_\mathrm{Ca}$ & 4.4 & 8 & 1 \\
    $\bar{g}_\mathrm{K}$  & 8 & 5 & 1 \\
    $\sigma_v$ & 0.0307 & 0.05 & 0.01 \\
    $\sigma_n$ & 0.001  & 0.01 & 0.001 \\
    $\sigma_{y,k}$ & 1  & 10 & 0.5 \\
  \hline
\end{tabular}
\caption{True value, initial value, and covariance of the parameters
in Fig. \ref{fig:PMCMC_res}.}\label{table:PMCMC_res_values}
\end{table}


\subsection{Estimation of synaptic conductances}
\label{sect:seq_estimation:results:syn}

Finally, once the methods to estimate state variables and unknown
parameters were consolidated, we proceeded to test the methods to
our ultimate goal: estimating jointly the intrinsic states of the
neuron and the extrinsic inputs (i.e., the synaptic conductances).

\begin{figure}[ht]
 \begin{center}
     \subfigure[]{\scalebox{1}{\includegraphics[width=8cm]{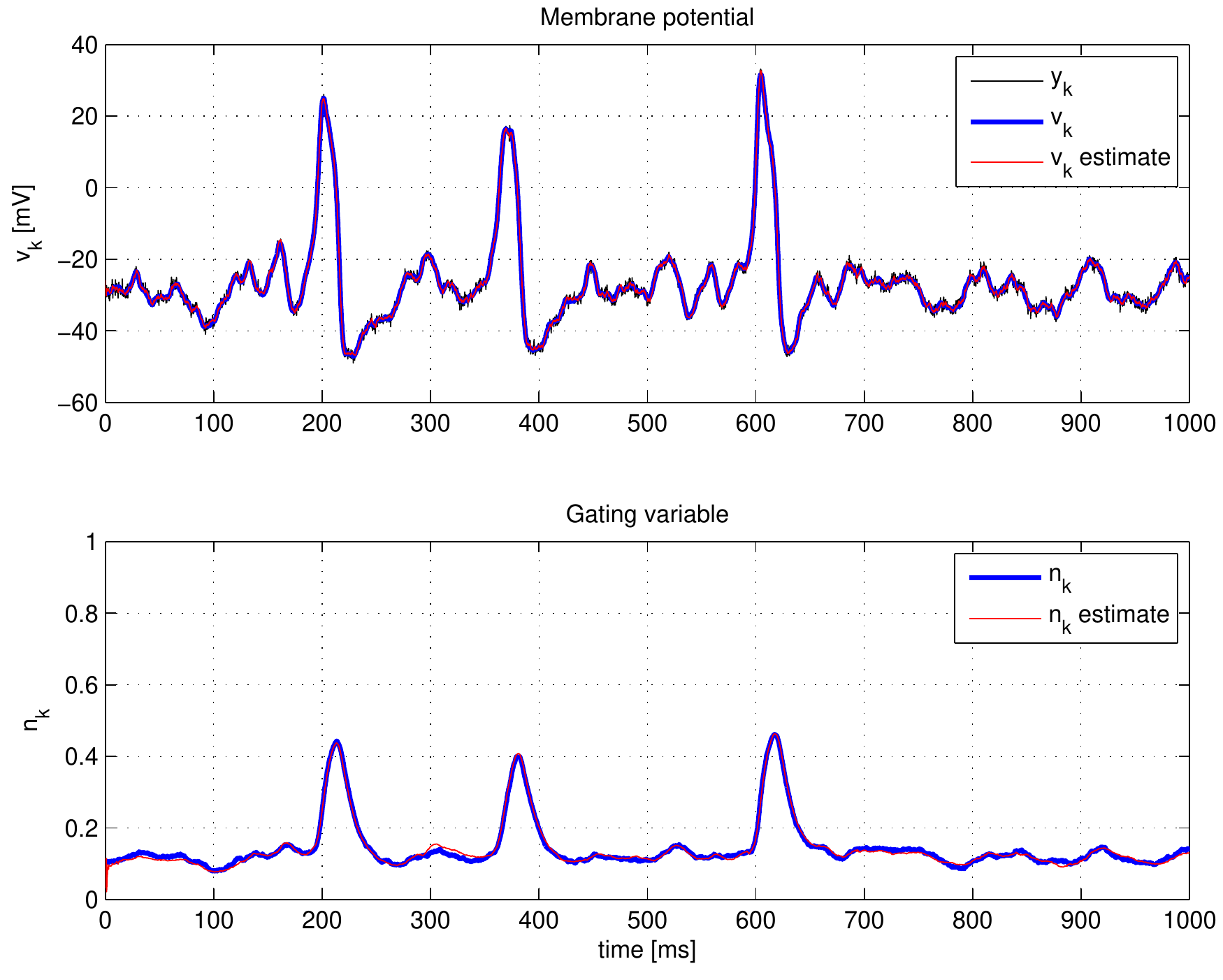}}\label{fig:ML_SNR32_N500_PF_SynCond}} \\
     \subfigure[]{\scalebox{1}{\includegraphics[width=8cm]{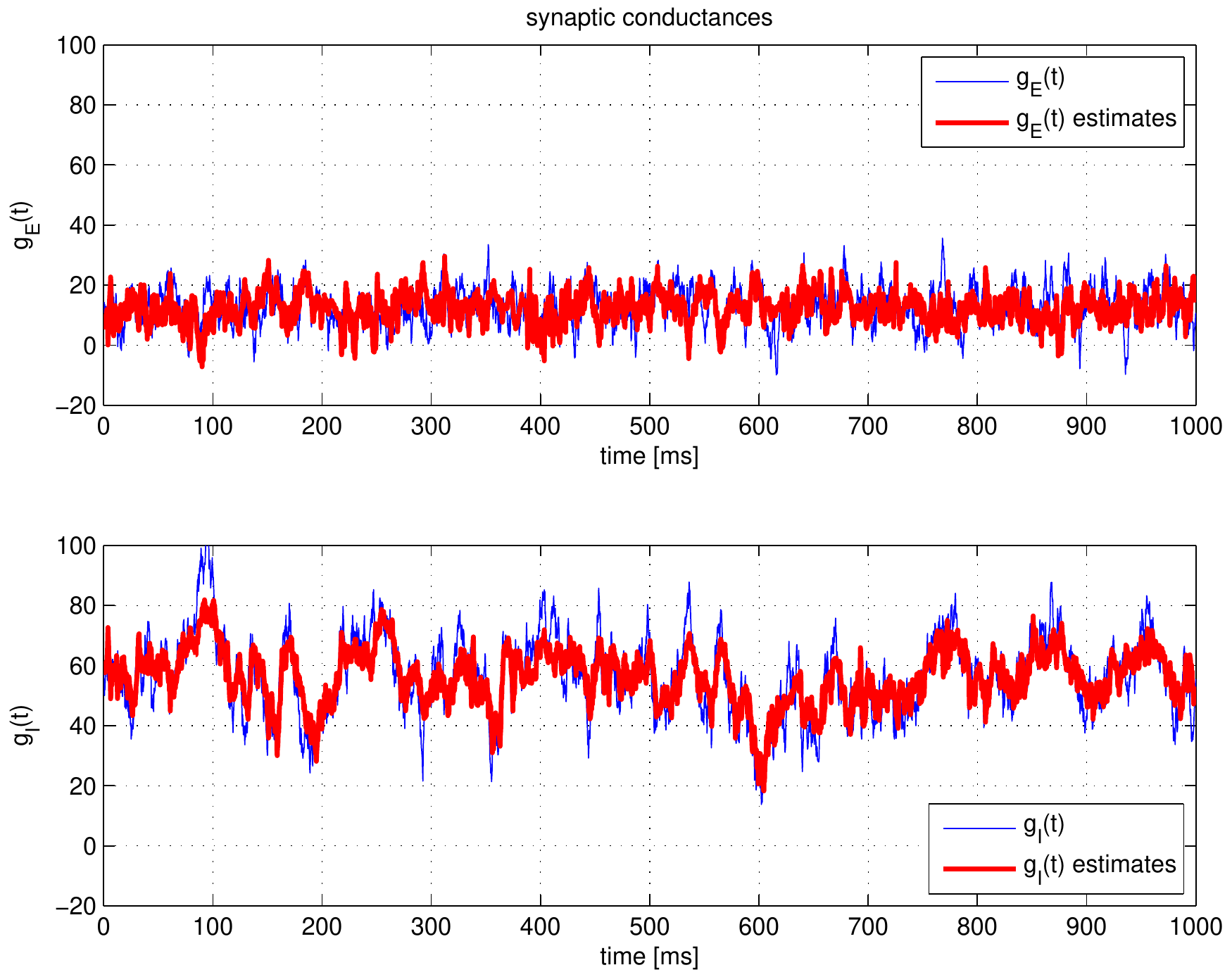}}\label{fig:ML_SNR32_N500_PF_SynCond2}}
  \caption{A single realization of the PF method with perfect model knowledge, estimating voltage and gating variables (top) and synaptic conductances in nS (bottom).}\label{fig:ML_SNR32_N500_PF_SynCond_all}
 \end{center}
\end{figure}

First, the method with perfect knowledge of the model was validated
in Fig. \ref{fig:ML_SNR32_N500_PF_SynCond_all}. It can be observed
in Fig. \ref{fig:ML_SNR32_N500_PF_SynCond} that the intrinsic
signals can be effectively recovered as before where synaptic inputs
were not accounted for. The estimation of $g_\mathrm{E}(t)$ and
$g_\mathrm{I}(t)$ is seen in Fig.
\ref{fig:ML_SNR32_N500_PF_SynCond2}. We see that the estimation of
the the excitatory and inhibitory terms is quite accurate, and that
the presence of spikes does not degrade the estimation capabilities
of the method.

\begin{figure}[ht]
 \begin{center}
     \subfigure[]{\scalebox{1}{\includegraphics[width=8cm]{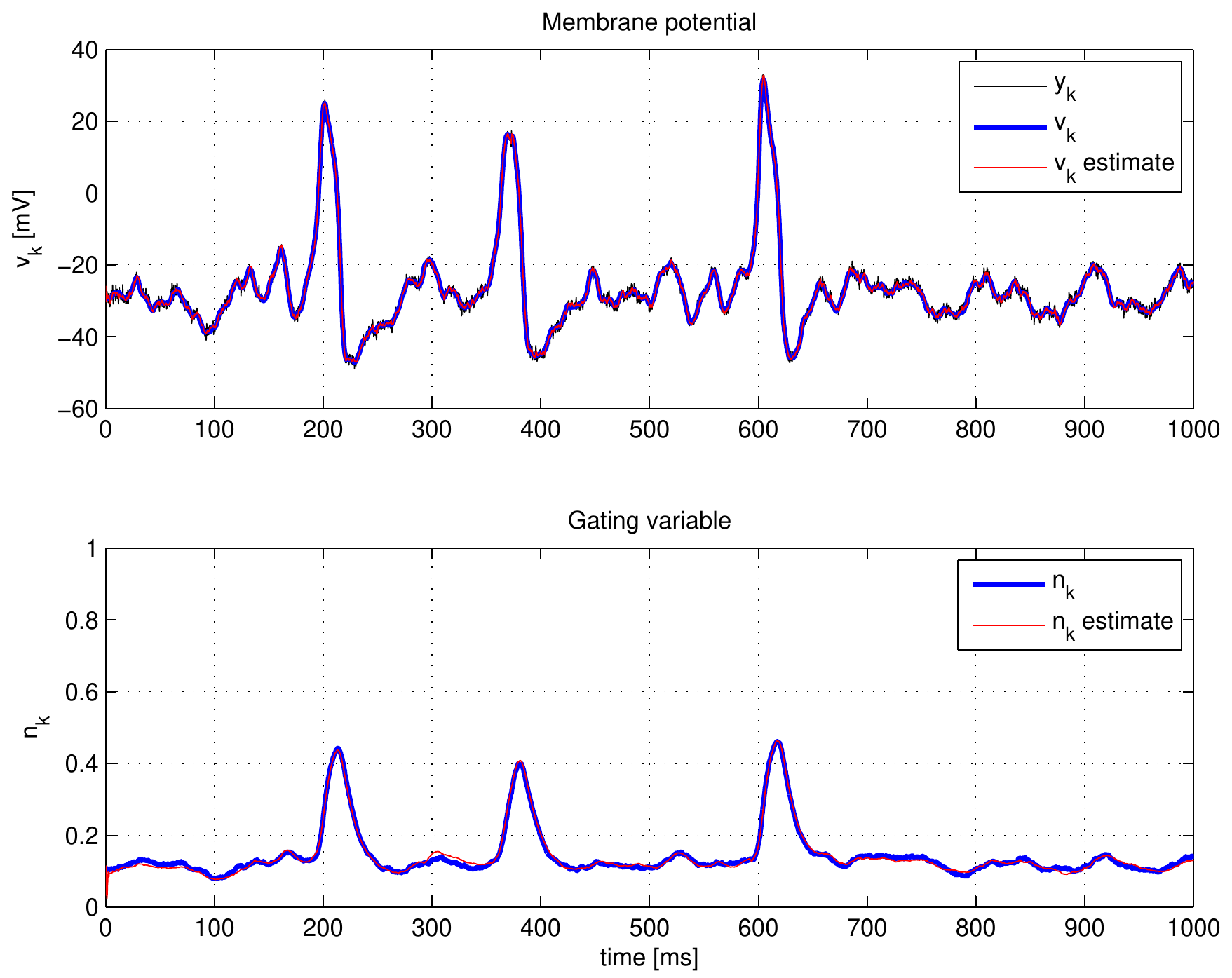}}\label{fig:ML_SNR32_N500_PMCMC_SynCond}} \\
     \subfigure[]{\scalebox{1}{\includegraphics[width=8cm]{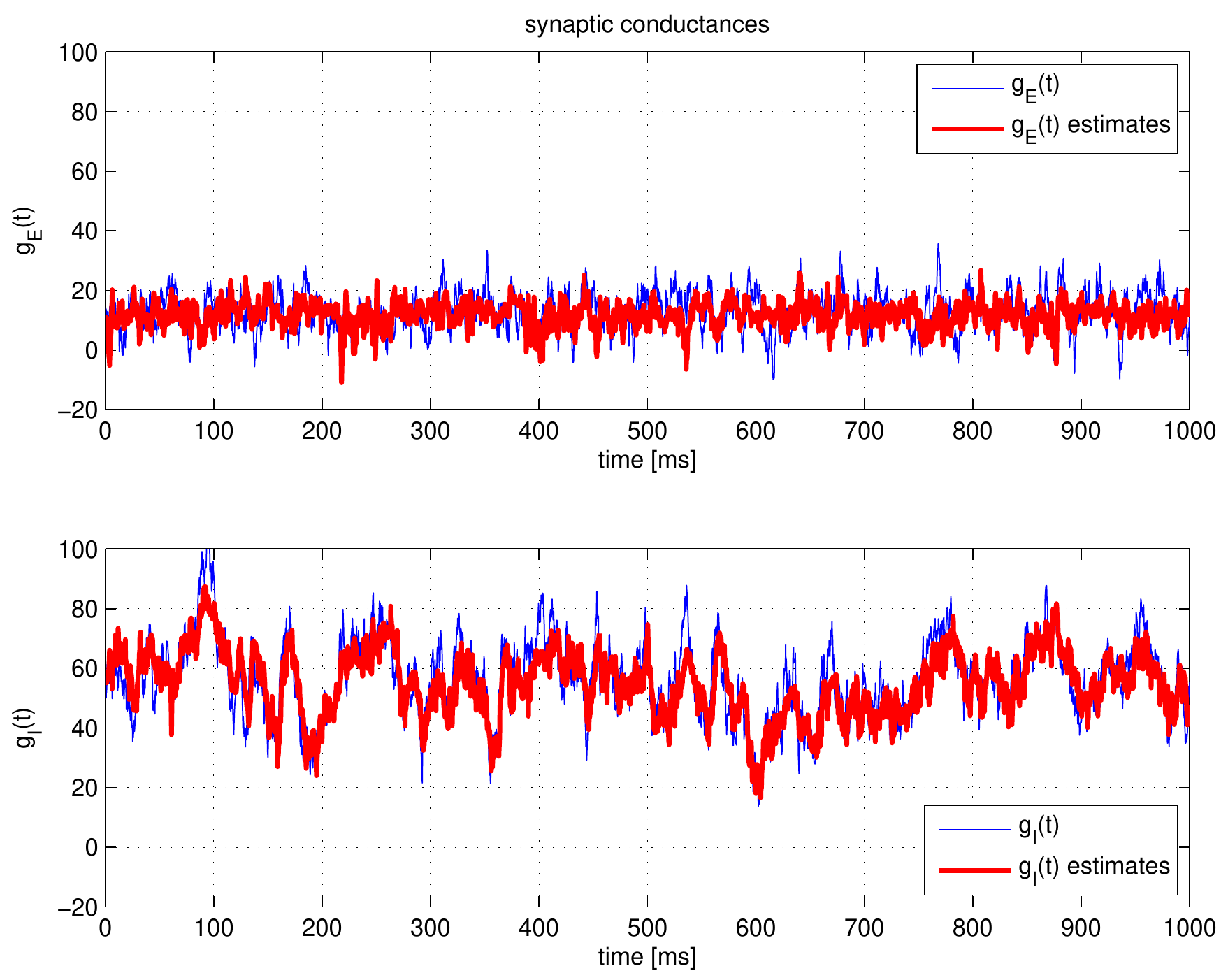}}\label{fig:ML_SNR32_N500_PMCMC_SynCond2}}
  \caption{A single realization of the PMCMC method, estimating voltage and gating variables (top) and synaptic conductances in nS (bottom) as well as model parameters.}\label{fig:ML_SNR32_N500_PMCMC_SynCond_all}
 \end{center}
\end{figure}

The \ac{PMCMC} algorithm was tested similarly. In this case, we
assumed that the model parameters related to $v_k$ and $n_k$ were
accurately estimated, for instance using an off-line procedure or
that analyzed in Section \ref{sect:seq_estimation:results:PFMCMC}.
Therefore, we focused on the estimation of those parameters that
describe the \ac{OU} process of each of the synaptic terms.
Particularly, we considered the values in Table
\ref{table:PMCMCsyn_res_values}. The results can be consulted in
Fig. \ref{fig:ML_SNR32_N500_PMCMC_SynCond_all} and compared to those
in Fig. \ref{fig:ML_SNR32_N500_PF_SynCond_all}. We observe that
little degradation with respect to the optimal case of perfectly
knowing the model can be identified.

\begin{table}[ht]
\centering
\begin{tabular}{|c|c|c|c|c|}
  \hline
    Parameter & True value & Initial value & Init. Covariance \\ \hline
    $\tau_\mathrm{E}$ & 2.73 & 1.5 & 1 \\
    $g_\mathrm{E,0}$  & 12.1 & 10 & 1 \\
    $\sigma_\mathrm{E}$ & 12 & 25 & 5 \\
    $\tau_\mathrm{I}$ & 10.49 & 15 & 10 \\
    $g_\mathrm{I,0}$  & 57.3 & 45 & 10 \\
    $\sigma_\mathrm{I}$  & 26.4 & 35 & 5 \\
  \hline
\end{tabular}
\caption{True value, initial value, and covariance of the parameters
in Fig.
\ref{fig:ML_SNR32_N500_PMCMC_SynCond_all}.}\label{table:PMCMCsyn_res_values}
\end{table}

We refer to the Supplementary Material to visualize a dynamic
simulation showing how the estimations evolve as the PMCMC algorithm
was applied in a case where the values of $\bar{g}_\mathrm{L}$ and
$E_\mathrm{L}$ were unknown.

\section{Conclusions}\label{sect:conclusions}

In this paper we propose a filtering method that is able to
sequentially infer the time-course of the membrane potential, the
intrinsic activity of ionic channels, and the input synaptic
conductances from noisy observations of voltage traces. The method works both for subthreshold and spiking regimes.
It is based on the \ac{PF} methodology and features an optimal
importance density, providing enhanced use of the particles that
characterize the filtering distribution. In addition, we tackle the
problem of joint parameter estimation and state filtering by
extending the designed \ac{PF} with an \ac{MCMC} procedure in an
iterative method known as \ac{PMCMC}. Another distinctive
contribution with respect to the other works in the literature is
that here we provide accuracy bounds for the problem at hand, given
by the \ac{PCRB}. The \acp{RMSE} of our methods are then compared to
the bound and, therefore, we can assess the efficiency of the
proposed inference methods.

Filtering methods of different types (e.g., \ac{PF} or sigma--point
Kalman filtering) have been used in other recent contributions to
similar problems, see \cite{Rudolph2004,Pospischil08,Bedard11,Kobayashi11,Paninski12,Berg2013,Lankarany2013a}. From a methodological perspective, the novelty of
this paper is in the use of an optimal importance density to
generate particles, fact that increases the estimation accuracy for
a given budget of particles. This technical detail only applies to
\ac{PF} methods. Although Gaussian methods (e.g. the family of
sigma--point Kalman filters) have a lower computational cost in
general, they require Gaussianity of the measures, whereas \acp{PF}
do not. This is an advantage that we think can be crucial in
estimating synaptic conductances, since the assumption of
Gaussianity is generally assumed in the literature
(\cite{Rudolph03,Lankarany2014}) but there are no conclusive
evidences to assert this assumption. In this paper, we have still
applied the \ac{PF} to an Ornstein-Uhlenbeck process in order to
check that basic results can be attained, but further research will
go in the direction of assuming other types of distributions for the
synaptic conductances. The use of more complex distributions nicely
fits within the framework of our \ac{PF}-based method. Another
advantage of PFs versus Gaussian filters is their enhanced
robustness to outliers \cite{Djuric09b}, for instance due to
recording artifacts, future applications shall also incorporate this
feature.

We have found excellent estimations of the synaptic conductances,
even in spiking regimes. Estimating synaptic conductances in spiking
regimes is a challenge which is far to be solved. It is well-known
that linear estimations of synaptic conductances are not trustable
in this situation when data is extracted intracellularly from
spiking activity of neurons, see \cite{Guillamon2006}. In
experiments, thus, caution has to be taken in eliminating part of
the voltage traces, thus loosing also part of the temporal
information of both excitatory and inhibitory conductances. Our
method is able to perform well in this regime. This information is
highly valuable in problems (epilepsy, schizophrenia, Alzheimer's disease,\dots) where a debate on the balance of
excitation and inhibition is open, see the introduction of
\cite{Berg2013} for a rather complete overview on this feature.

The results show the validity of the approach when applied to
Morris-Lecar type of neuron. However, the procedure is general and
could be applied to any neuron model, exhibiting more complex dynamics like bursting, mixed-mode oscillations,\dots Forthcoming applications would
be validating the method using real data recordings, both for
inferring parameters of the model and synaptic conductances. The
latter problem is a challenging \emph{hot} topic in the neuroscience
literature, which is recently focusing on methods to extract the
conductances from single-trace measurements. We think that our
\ac{PF} method would give useful and interesting results to
physiologists that aim at inferring brain's activation rules from
neurons' activities. Actually, knowing the excitatory-inhibitory
time course separation  can help in getting important conclusions
about brain's functional connectivity (see
\cite{Anderson2000,WZ2003,Bennet2013}).

We have not tried to obtain estimations when
subthreshold ionic currents are active, where the presence of nonlinearities
could also contaminate the estimations, see \cite{Vich2015}.
According to the excellent performance in spiking regimes, where
nonlinearities are stronger, we expect also a good agreement between
the estimated data and the prescribed synaptic conductances. Other
extensions of the model can be devoted to incorporate the
dendro-somatic interaction (see for instance \cite{Cox2004}), by
considering multi-compartmental neuron models, thus taking into account the morphology and the functional properties of the cell. This is another big
challenge  for which we think that our method can account for.

\appendices
\section{Morris-Lecar neuron model}\label{ap:MLmodel}

From the myriad of existing single-neuron models, we consider
without loss of generality the Morris-Lecar model proposed in
\cite{Morris81}. The model can be related (see \cite{Izhikevich06})
to the $I_\mathrm{Na,p}+I_\mathrm{K}$-model (pronounced persistent
sodium plus potassium model). The dynamics of the neuron is modeled
by a continuous-time dynamical system composed of the
current-balance equation for the membrane potential, $v=v(t)$, and
the K$^+$ gating variable $0\leq n = n(t) \leq 1$, which represents
the probability of the K$^+$ ionic channel to be active. Then, the
system of differential equations is
\begin{eqnarray}
C_m \dot{v} &=& - I_\mathrm{L} - I_\mathrm{Ca} - I_\mathrm{K} + I_{\mathrm{app}} \label{eq:ML_ODE1} \\
\dot{n} &=& \phi \frac{n_\infty(v)-n}{\tau_{n}(v)}
\label{eq:ML_ODE2} ~,
\end{eqnarray}
\noindent where $C_m$ is the membrane capacitance and $\phi$ a
non-dimensional constant. $I_{\mathrm{app}}$ represents the
(externally) applied current. For the time being, we have neglected
$I_{\mathrm{syn}}$ in (\ref{eq:ML_ODE1}). The leakage, calcium, and
potassium currents are of the form
\begin{eqnarray}
  I_\mathrm{L}  &=& \bar{g}_\mathrm{L} (v-E_\mathrm{L}) \\
  I_\mathrm{Ca} &=& \bar{g}_\mathrm{Ca} m_\infty(v) (v-E_\mathrm{Ca}) \\
  I_\mathrm{K}  &=& \bar{g}_\mathrm{K} n (v-E_\mathrm{K})~,
\end{eqnarray}
\noindent respectively. $\bar{g}_\mathrm{L}$, $\bar{g}_\mathrm{Ca}$,
and $\bar{g}_\mathrm{K}$ are the maximal conductances of each
current. $E_\mathrm{L}$, $E_\mathrm{Ca}$, and $E_\mathrm{K}$ denote
the Nernst equilibrium potentials, for which the corresponding
current is zero, a.k.a. reverse potentials.

The dynamics of the activation variable $m$ is considered at the
steady state, and thus we write $m=m_\infty(v)$. On the other hand,
the time constant $\tau_n(v)$ for the gating variable $n$ cannot be
considered that fast and the corresponding differential equation
needs to be considered. The formulae for these functions is
\begin{eqnarray}\label{eq:sigmoids}
m_\infty(v) &=& \frac{1}{2} \cdot (1 + \tanh [\frac{v-V_1}{V_2} ]) \label{eq:sigmoids1} \\
n_\infty(v) &=& \frac{1}{2} \cdot (1 + \tanh [\frac{v-V_3}{V_4} ]) \label{eq:sigmoids2} \\
\tau_n(v) &=& 1 / ( \cosh [\frac{v-V_3}{2V_4} ] )
\label{eq:sigmoids3} ~,
\end{eqnarray}
\noindent which parameters $V_1$, $V_2$, $V_3$, and $V_4$ can be
measured experimentally \cite{Izhikevich06}.

The knowledgeable reader would have noticed that the Morris-Lecar
model is a Hodgin-Huxley type-model with the usual considerations,
where the following two extra assumptions were made: the
depolarizing current is generated by Ca$^{2+}$ ionic channels (or
Na$^+$ depending on the type of neuron modeled), whereas
hyperpolarization is carried by K$^+$ ions; and that 
$m=m_\infty(v)$. The Morris-Lecar model is very popular in
computational neuroscience as it models a large variety of neural
dynamics while its phase-plane analysis is more manageable as it
involves only two states \cite{Rinzel98}.

The Morris-Lecar, although simple to formulate, results in a very
interesting model as it can produce a number of different dynamics.
For instance, for given values of its parameters, we encounter a
subcritic Hopf bifurcation for $I_{\mathrm{app}}=93.86$
$\mu$A/cm$^2$. On the other hand, for another set of parameter
values, the system of equations has a Saddle-Node on an Invariant
Circle (SNIC) bifurcation at $I_{\mathrm{app}}=39.96$ $\mu$A/cm$^2$.


\section{PCRB in Morris-Lecar models}\label{ap:seq_estimation:bounds}

This appendix is devoted to the derivation of the \ac{PCRB}
estimation bound for the Morris-Lecar model used to benchmark the
proposed methods in the simulations. We follow the sequential
procedure given in \cite{Tichavsky98}, accounting that we have
nonlinear functions in the state evolution and linear measurements,
both with additive Gaussian noise. We are interested in an
estimation error bound of the type of
\begin{equation}\label{eq:PCRB_general}
\mathbb{E}_{y_k,\bfx_k} \left\{ (\hat{\bfx}_k(y_{1:k})-\bfx_k)
(\hat{\bfx}_k(y_{1:k})-\bfx_k)^\top \right\} \geq \mathbf{J}_k^{-1}
~,
\end{equation}
\noindent where $\hat{\bfx}_k(y_{1:k})$ represents an estimator of
$\bfx_k$ given $y_{1:k}$.

Recall that the state-space we are dealing with is of the form
\begin{eqnarray}
\nonumber \bfx_k &=& \mathbf{f}_{k-1}(\bfx_{k-1}) + \bm{\nu}_{k} \\
y_k &=& \mathbf{h} \bfx_{k} + e_{k} ~, 
\end{eqnarray}
\noindent where $\mathbf{h} = (1,0)$, $\bfx_k = \left(v_k , n_k
\right)^\top$, and $\mathbf{f}_{k-1}(\bfx_{k-1})$ defined by
(\ref{eq:ML_discrete1}) and (\ref{eq:ML_discrete2}). The noise terms
are of the form
\begin{eqnarray}
    \bm{\nu}_k & \sim & \mathcal{N}(\mathbf{0},\bm{\Sigma}_{x,k}) \\
    e_k & \sim & \mathcal{N} (0,\sigma_{y,k}^2) ~.
\end{eqnarray}
In this case, the \ac{PCRB} can be computed recursively by virtue of
the result in \cite{Tichavsky98} by computing the following terms
\begin{eqnarray}
\mathbf{D}_k^{11} &=& \mathbb{E}_{\bfx_{k}}  \left\{\tilde{\mathbf{F}}_k^\top \bm{\Sigma}_{x,k}^{-1} \tilde{\mathbf{F}}_k \right\} \\
\mathbf{D}_k^{12} &=& \mathbf{D}_k^{21} = - \mathbb{E}_{\bfx_{k}}  \left\{\tilde{\mathbf{F}}_k^\top \right\} \bm{\Sigma}_{x,k}^{-1} \\
\mathbf{D}_k^{22} &=& \bm{\Sigma}_{x,k}^{-1} + \mathbf{H}_{k+1}^\top
\bm{\Sigma}_{y,k+1}^{-1} \mathbf{H}_{k+1}
\end{eqnarray}
\noindent and plugging them into
\begin{equation}\label{eq:PCRB_recursive}
\mathbf{J}_{k+1} = \mathbf{D}_k^{22} - \mathbf{D}_k^{21} \left(
\mathbf{J}_k + \mathbf{D}_k^{11} \right)^{-1} \mathbf{D}_k^{12} ~,
\end{equation}
\noindent for some initial $\mathbf{J}_0$. Notice that, in our case,
$\mathbf{D}_k^{22}$ becomes deterministic, but the rest of terms
involving expectations should be computed by Monte Carlo integration
over independent state trajectories.

Since the state function is nonlinear, we use the Jacobian evaluated
at the true value of $\bfx_k$ instead, that is
\begin{eqnarray}
\tilde{\mathbf{F}}_{k} &=& \left[ \nabla_{\bfx_{k}}
\mathbf{f}_{k}^\top ( \bfx_{k} ) \right]^\top = \left(
\begin{array}{cc}
           \frac{\partial f_1}{\partial v_t} & \frac{\partial f_1}{\partial n_t} \\
           \frac{\partial f_2}{\partial v_t} & \frac{\partial f_2}{\partial n_t}
         \end{array}
\right)~,
\end{eqnarray}
\noindent where functions $f_1$ and $f_2$ are as in
(\ref{eq:ML_discrete1}) and (\ref{eq:ML_discrete2}), respectively.
Therefore, to evaluate the bound we need to compute the derivatives
in the Jacobian. These are,
\begin{multline*}
    \frac{\partial f_1( \bfx_{k} )}{\partial v_k} =  \\
    \nonumber 1 - \frac{\Ts}{C_m} \left(\bar{g}_\mathrm{L} + \bar{g}_\mathrm{K} n_k + \bar{g}_\mathrm{Ca} \frac{\partial m_\infty(v_{k})}{\partial v_k} v_k + \bar{g}_\mathrm{Ca} m_\infty(v_{k})  \right)
\end{multline*}
\begin{multline*}
    \frac{\partial f_2( \bfx_{k} )}{\partial v_k} = \\
    \nonumber \Ts \phi \frac{\frac{\partial n_\infty(v_{k})}{\partial v_k}\tau_{n}(v_{k}) - (n_\infty(v_{k-1})-n_{k-1}) \frac{\partial \tau_{n}(v_{k})}{\partial v_k}}{\tau_{n}^2(v_{k})}
\end{multline*}
\begin{multline*}
    \frac{\partial f_1( \bfx_{k} )}{\partial n_k} = - \frac{\Ts}{C_m} \bar{g}_\mathrm{K} (v_{k}-E_\mathrm{K})
\end{multline*}
\begin{multline*}
    \frac{\partial f_2( \bfx_{k} )}{\partial n_k} = 1- \frac{\Ts \phi}{\tau_{n}(v_{k})}
\end{multline*}
\noindent with
\begin{eqnarray}
    \frac{\partial m_\infty(v_{k})}{\partial v_k} &=&  \frac{1}{2 V_2} \textrm{sech}^2\left( \frac{v_k - V_1}{V_2} \right) \\
    \frac{\partial n_\infty(v_{k})}{\partial v_k} &=&  \frac{1}{2 V_4} \textrm{sech}^2\left( \frac{v_k - V_3}{V_4} \right) \\
    \frac{\partial \tau_{n}(v_{k})}{\partial v_k} &=& -\frac{1}{2 V_4} \frac{\textrm{sinh}\left( \frac{v_k - V_3}{2V_4} \right)}{\textrm{cosh}^2\left( \frac{v_k - V_3}{2V_4} \right)} ~.
\end{eqnarray}

\bibliographystyle{IEEEtran}
\bibliography{bibliografia,pau,IEEEabrv}   

\begin{thebibliography}{10}
\providecommand{\url}[1]{#1}
\csname url@samestyle\endcsname
\providecommand{\newblock}{\relax}
\providecommand{\bibinfo}[2]{#2}
\providecommand{\BIBentrySTDinterwordspacing}{\spaceskip=0pt\relax}
\providecommand{\BIBentryALTinterwordstretchfactor}{4}
\providecommand{\BIBentryALTinterwordspacing}{\spaceskip=\fontdimen2\font plus
\BIBentryALTinterwordstretchfactor\fontdimen3\font minus
  \fontdimen4\font\relax}
\providecommand{\BIBforeignlanguage}[2]{{%
\expandafter\ifx\csname l@#1\endcsname\relax
\typeout{** WARNING: IEEEtran.bst: No hyphenation pattern has been}%
\typeout{** loaded for the language `#1'. Using the pattern for}%
\typeout{** the default language instead.}%
\else
\language=\csname l@#1\endcsname
\fi
#2}}
\providecommand{\BIBdecl}{\relax}
\BIBdecl

\bibitem{Brette12}
\BIBentryALTinterwordspacing
R.~Brette and A.~Destexhe, \emph{Handbook of Neural Activity
  Measurement}.\hskip 1em plus 0.5em minus 0.4em\relax Cambridge University
  Press, 2012. [Online]. Available:
  \url{http://dx.doi.org/10.1017/CBO9780511979958}
\BIBentrySTDinterwordspacing

\bibitem{Huys09}
Q.~J. Huys and L.~Paninski, ``Smoothing of, and parameter estimation from,
  noisy biophysical recordings,'' \emph{PLoS Computational Biology}, vol.~5,
  no.~5, pp. 1--16, 2009.

\bibitem{DitlevsenSamson2014}
\BIBentryALTinterwordspacing
S.~Ditlevsen and A.~Samson, ``{Estimation in the partially observed stochastic
  Morris-Lecar neuronal model with particle filter and stochastic approximation
  methods},'' \emph{Ann. Appl. Stat.}, vol.~8, no.~2, pp. 674--702, 06 2014.
  [Online]. Available: \url{http://dx.doi.org/10.1214/14-AOAS729}
\BIBentrySTDinterwordspacing

\bibitem{Lankarany2014}
\BIBentryALTinterwordspacing
M.~Lankarany, W.-P. Zhu, and M.~Swamy, ``{Joint estimation of states and
  parameters of Hodgkin-Huxley neuronal model using Kalman filtering},''
  \emph{Neurocomputing}, vol. 136, pp. 289 -- 299, 2014. [Online]. Available:
  \url{http://www.sciencedirect.com/science/article/pii/S0925231214001155}
\BIBentrySTDinterwordspacing

\bibitem{DitlevsenSamson2015}
S.~Ditlevsen and A.~Samson, ``Parameter estimation in neuronal stochastic
  differential equation models from intracellular recordings of membrane
  potentials in single neurons: a review,'' \emph{Journal de la Soci{\'e}t{\'e}
  Fran{\c{c}}aise de Statistique}, vol. to appear, 2015.

\bibitem{Mishchenko2011}
\BIBentryALTinterwordspacing
Y.~Mishchenko, J.~T. Vogelstein, and L.~Paninski, ``{A Bayesian approach for
  inferring neuronal connectivity from calcium fluorescent imaging data},''
  \emph{Ann. Appl. Stat.}, vol.~5, no.~2B, pp. 1229--1261, 06 2011. [Online].
  Available: \url{http://dx.doi.org/10.1214/09-AOAS303}
\BIBentrySTDinterwordspacing

\bibitem{Rudolph2004}
\BIBentryALTinterwordspacing
M.~Rudolph, Z.~Piwkowska, M.~Badoual, T.~Bal, and A.~Destexhe, ``A method to
  estimate synaptic conductances from membrane potential fluctuations,''
  \emph{Journal of Neurophysiology}, vol.~91, no.~6, pp. 2884--2896, 2004.
  [Online]. Available: \url{http://jn.physiology.org/content/91/6/2884.full}
\BIBentrySTDinterwordspacing

\bibitem{Pospischil08}
M.~Pospischil, M.~Toledo-Rodriguez, C.~Monier, Z.~Piwkowska, T.~Bal,
  Y.~Frégnac, H.~Markram, and A.~Destexhe, ``{Minimal Hodgkin-Huxley type
  models for different classes of cortical and thalamic neurons},''
  \emph{Biological Cybernetics}, vol.~99, no. 4-5, pp. 427--441, 2008.

\bibitem{Bedard11}
C.~B\'{e}dard, S.~B\'{e}huret, C.~Deleuze, T.~Bal, and A.~Destexhe,
  ``{Oversampling method to extract excitatory and inhibitory conductances from
  single-trial membrane potential recordings.}'' \emph{Journal of neuroscience
  methods}, Sep. 2011.

\bibitem{Kobayashi11}
R.~Kobayashi, Y.~Tsubo, P.~Lansky, and S.~Shinomoto, ``Estimating time-varying
  input signals and ion channel states from a single voltage trace of a
  neuron,'' \emph{Advances in Neural Information Processing Systems (NIPS)},
  vol.~24, pp. 217--225, 2011.

\bibitem{Paninski12}
L.~Paninski, M.~Vidne, B.~DePasquale, and D.~G. Ferreira, ``{Inferring synaptic
  inputs given a noisy voltage trace via sequential Monte Carlo methods},''
  \emph{Journal of Computational Neuroscience}, vol.~33, no.~1, pp. 1--19,
  2012.

\bibitem{Berg2013}
R.~W. Berg and S.~Ditlevsen, ``Synaptic inhibition and excitation estimated via
  the time constant of membrane potential fluctuations,'' \emph{Journal of
  Neurophysiology}, vol. 110, no.~4, pp. 1021--1034, 2013.

\bibitem{Lankarany2013a}
\BIBentryALTinterwordspacing
M.~Lankarany, W.~P. Zhu, M.~N.~S. Swamy, and T.~Toyoizumi, ``{Inferring
  trial-to-trial excitatory and inhibitory synaptic inputs from membrane
  potential using Gaussian mixture Kalman filtering},'' \emph{Frontiers in
  Computational Neuroscience}, vol.~7, 2013. [Online]. Available:
  \url{http://dx.doi.org/10.3389/fncom.2013.00109}
\BIBentrySTDinterwordspacing

\bibitem{Guillamon2006}
\BIBentryALTinterwordspacing
A.~Guillamon, D.~W. McLaughlin, and J.~Rinzel, ``{Estimation of synaptic
  conductances},'' \emph{Journal of Physiology-Paris}, vol. 100, no. 1-3, pp.
  31--42, Jul. 2006. [Online]. Available:
  \url{http://dx.doi.org/10.1016/j.jphysparis.2006.09.010}
\BIBentrySTDinterwordspacing

\bibitem{Vich2015}
\BIBentryALTinterwordspacing
C.~Vich and A.~Guillamon, ``\BIBforeignlanguage{English}{Dissecting estimation
  of conductances in subthreshold regimes},''
  \emph{\BIBforeignlanguage{English}{Journal of Computational Neuroscience}},
  pp. 1--17, 2015. [Online]. Available:
  \url{http://dx.doi.org/10.1007/s10827-015-0576-2}
\BIBentrySTDinterwordspacing

\bibitem{Closas13}
P.~Closas and A.~Guillamon, ``{Sequential estimation of gating variables from
  voltage traces in single-neuron models by particle filtering},'' in
  \emph{Proceedings of the IEEE International Conference on Acoustics, Speech,
  and Signal Processing, ICASSP 2013}, Vancouver, Canada, May 2013.

\bibitem{Closas13b}
------, ``Estimation of neural voltage traces and associated variables in
  uncertain models,'' \emph{BMC Neuroscience}, vol.~14, no.~1, p. 1151, July
  2013.

\bibitem{SeqMC01}
A.~Doucet, N.~de~Freitas, and N.~Gordon, Eds., \emph{Sequential Monte Carlo
  Methods in Practice}.\hskip 1em plus 0.5em minus 0.4em\relax Springer, 2001.

\bibitem{SpecialIssue02}
``{Special Issue on Monte Carlo methods for Statistical Signal Processing},''
  vol.~50, no.~2, February 2002.

\bibitem{DjuricM03}
P.~M. Djuri\'c, J.~H. Kotecha, J.~Zhang, Y.~Huang, T.~Ghirmai, M.~F. Bugallo,
  and J.~M\'iguez, ``{Particle Filtering},'' vol.~20, no.~5, pp. 19--38,
  September 2003.

\bibitem{Arulampalam02}
S.~Arulampalam, S.~Maskell, N.~Gordon, and T.~Clapp, ``{A Tutorial on Particle
  Filters for Online Nonlinear/Non-Gaussian Bayesian Tracking},'' vol.~50,
  no.~2, pp. 174--188, February 2002.

\bibitem{Chen03}
Z.~Chen, ``{Bayesian filtering: From Kalman filters to particle filters, and
  beyond},'' Adaptive Syst. Lab., McMaster University, Ontario, Canada, Tech.
  Rep., 2003.

\bibitem{BeyondKF}
B.~Ristic, S.~Arulampalam, and N.~Gordon, \emph{Beyond the Kalman Filter:
  Particle Filters for tracking applications}.\hskip 1em plus 0.5em minus
  0.4em\relax Boston: Artech House, 2004.

\bibitem{Sarkka13book}
S.~S\"arkk\"a, \emph{Bayesian Filtering and Smoothing}.\hskip 1em plus 0.5em
  minus 0.4em\relax New York, NY, USA: Cambridge University Press, 2013.

\bibitem{Doucet00}
A.~Doucet, S.~J. Godsill, and C.~Andrieu, ``{On sequential Monte Carlo sampling
  methods for Bayesian filtering},'' \emph{Stat. Comput.}, vol.~3, pp.
  197--208, 2000.

\bibitem{Ullah09}
G.~Ullah and S.~J. Schiff, ``{Tracking and control of neuronal Hodgkin-Huxley
  dynamics},'' \emph{Physical Review E}, vol.~79, no.~4, p. 040901, 2009.

\bibitem{Dayan05}
P.~Dayan and L.~F. Abbott, \emph{Theoretical Neuroscience: Computational and
  Mathematical Modeling of Neural Systems}.\hskip 1em plus 0.5em minus
  0.4em\relax The MIT Press, 2005.

\bibitem{Izhikevich06}
E.~Izhikevich, \emph{{Dynamical systems in neuroscience: the geometry of
  excitability and bursting}}.\hskip 1em plus 0.5em minus 0.4em\relax
  Cambridge, MA: MIT Press, 2006.

\bibitem{Keener09}
\BIBentryALTinterwordspacing
J.~P. Keener and J.~Sneyd, \emph{Mathematical physiology. {I}. , Cellular
  physiology}, ser. Interdisciplinary applied mathematics.\hskip 1em plus 0.5em
  minus 0.4em\relax New York, London: Springer, 2009. [Online]. Available:
  \url{http://opac.inria.fr/record=b1133357}
\BIBentrySTDinterwordspacing

\bibitem{ErmentroutTerman2010}
\BIBentryALTinterwordspacing
B.~Ermentrout and D.~H. Terman, \emph{Mathematical foundations of
  neuroscience}, ser. Interdisciplinary applied mathematics.\hskip 1em plus
  0.5em minus 0.4em\relax New York, Dordrecht, Heidelberg: Springer, 2010.
  [Online]. Available: \url{http://www.springer.com/us/book/9780387877075}
\BIBentrySTDinterwordspacing

\bibitem{Rudolph03}
M.~Rudolph and A.~Destexhe, ``Characterization of subthreshold voltage
  fluctuations in neuronal membranes,'' \emph{Neural Comput.}, vol.~15, no.~11,
  pp. 2577--2618, Nov. 2003.

\bibitem{Hodgkin52}
A.~L. Hodgkin and A.~F. Huxley, ``{The components of membrane conductance in
  the giant axon of Loligo},'' \emph{J Physiol.}, vol. 116, no.~4, pp.
  473--496, April 1952.

\bibitem{FitzHugh61}
R.~FitzHugh, ``Impulses and physiological states in theoretical models of nerve
  membrane,'' \emph{Biophys. J.}, vol.~1, pp. 445--466, 1961.

\bibitem{Nagumo62}
J.~Nagumo, S.~Arimoto, and S.~Yoshizawa, ``An active pulse transmission line
  simulating nerve axon,'' \emph{Proceeding IRE}, vol.~50, pp. 2061--2070,
  1962.

\bibitem{Morris81}
C.~Morris and H.~Lecar, ``{Voltage Oscillations in the barnacle giant muscle
  fiber},'' \emph{Biophys J.}, vol.~35, no.~1, pp. 193--213, July 1981.

\bibitem{Izhikevich04}
E.~Izhikevich, ``Which model to use for cortical spiking neurons?''
  \emph{Neural Networks, IEEE Transactions on}, vol.~15, no.~5, pp. 1063--1070,
  Sept 2004.

\bibitem{Rabinovich06}
M.~I. Rabinovich, P.~Varona, A.~I. Selverston, and H.~D. Abarbanel, ``Dynamical
  principles in neuroscience,'' \emph{Reviews of modern physics}, vol.~78,
  no.~4, p. 1213, 2006.

\bibitem{Douc05}
R.~Douc, O.~Capp\'{e}, and E.~Moulines, ``Comparison of resampling schemes for
  particle filtering,'' in \emph{Proc. of the 4th International Symposium on
  Image and Signal Processing and Analysis, ISPA'05}, Zagreb, Croatia, Sept.
  2005, pp. 64--69.

\bibitem{Andrieu04}
C.~Andrieu, A.~Doucet, S.~Singh, and V.~Tadic, ``Particle methods for change
  detection, system identification, and control,'' \emph{Proceedings of the
  IEEE}, vol.~92, no.~3, pp. 423--438, Mar 2004.

\bibitem{Andrieu05}
C.~Andrieu, A.~Doucet, and V.~B. Tadic, ``On-line parameter estimation in
  general state-space models,'' in \emph{Decision and Control, 2005 and 2005
  European Control Conference. CDC-ECC '05. 44th IEEE Conference on}, Dec 2005,
  pp. 332--337.

\bibitem{Poyiadjis11}
G.~Poyiadjis, A.~Doucet, and S.~S. Singh, ``{Particle approximations of the
  score and observed information matrix in state space models with application
  to parameter estimation},'' \emph{Biometrika}, vol.~98, pp. 65--80, 2011.

\bibitem{Andrieu10}
C.~Andrieu, A.~Doucet, and R.~Holenstein, ``{Particle Markov Chain Monte Carlo
  methods},'' \emph{Journal of the Royal Statistical Society Series B},
  vol.~72, no.~3, pp. 269--342, 2010.

\bibitem{Gilks96}
W.~Gilks, S.~Richardson, and D.~Spiegelhalter, Eds., \emph{Markov Chain Monte
  Carlo in Practice: Interdisciplinary Statistics}, ser. CRC Interdisciplinary
  Statistics Series.\hskip 1em plus 0.5em minus 0.4em\relax Chapman \& Hall,
  1996.

\bibitem{Berzuini97}
C.~Berzuini, N.~Best, W.~Gilks, and C.~Larizza, ``{Dynamic conditional
  independence models and Markov Chain Monte Carlo methods},'' \emph{Journal of
  American Stat. Assoc.}, vol.~92, pp. 1403--1412, 1997.

\bibitem{Liu05}
J.~S. Liu, \emph{{Monte Carlo Strategies in Scientific Computing}}.\hskip 1em
  plus 0.5em minus 0.4em\relax Springer, Jan. 2008.

\bibitem{Brooks11}
S.~Brooks, A.~Gelman, G.~Jones, and X.-L. Meng, \emph{Handbook of Markov Chain
  Monte Carlo}.\hskip 1em plus 0.5em minus 0.4em\relax CRC press, 2011.

\bibitem{Donnet2014}
S.~Donnet and A.~Samson, ``{Using PMCMC in EM algorithm for stochastic mixed
  models: theoretical and practical issues},'' \emph{Journal de la
  Soci{\'e}t{\'e} Fran{\c{c}}aise de Statistique}, vol. 155, no.~1, pp. 49--72,
  2014.

\bibitem{Vihola12}
M.~Vihola, ``Robust adaptive metropolis algorithm with coerced acceptance
  rate.'' \emph{Statistics and Computing}, vol.~22, no.~5, pp. 997--1008, 2012.

\bibitem{Golub96}
G.~H. Golub and C.~F. van Loan, \emph{Matrix Computations}, 3rd~ed.\hskip 1em
  plus 0.5em minus 0.4em\relax The John Hopkins University Press, 1996.

\bibitem{VanTrees07}
H.~L.~V. Trees and K.~L. Bell, Eds., \emph{Bayesian Bounds for Parameter
  Estimation and Nonlinear Filtering/Tracking}.\hskip 1em plus 0.5em minus
  0.4em\relax Wiley Interscience, 2007.

\bibitem{Djuric09b}
P.~M. Djuri\'c, M.~F. Bugallo, P.~Closas, and J.~M\'iguez, ``{Measuring the
  robustness of sequential methods},'' in \emph{Proceedings of the IEEE
  International Workshop on Computational Advances in Multi-Sensor Adaptive
  Processing, CAMSAP'09}, Aruba (Dutch Antilles), December 2009.

\bibitem{Anderson2000}
\BIBentryALTinterwordspacing
J.~S. Anderson, M.~Carandini, and D.~Ferster, ``{Orientation tuning of input
  conductance, excitation, and inhibition in cat primary visual cortex.}''
  \emph{Journal of neurophysiology}, vol.~84, no.~2, pp. 909--926, Aug. 2000.
  [Online]. Available: \url{http://jn.physiology.org/content/84/2/909.abstract}
\BIBentrySTDinterwordspacing

\bibitem{WZ2003}
\BIBentryALTinterwordspacing
M.~Wehr and A.~M. Zador, ``{Balanced inhibition underlies tuning and sharpens
  spike timing in auditory cortex.}'' \emph{Nature}, vol. 426, no. 6965, pp.
  442--446, Nov. 2003. [Online]. Available:
  \url{http://dx.doi.org/10.1038/nature02116}
\BIBentrySTDinterwordspacing

\bibitem{Bennet2013}
\BIBentryALTinterwordspacing
C.~Bennett, S.~Arroyo, and S.~Hestrin, ``Subthreshold mechanisms underlying
  state-dependent modulation of visual responses,'' \emph{Neuron}, vol.~80,
  no.~2, pp. 350 -- 357, 2013. [Online]. Available:
  \url{http://www.sciencedirect.com/science/article/ pii/S0896627313007186}
\BIBentrySTDinterwordspacing

\bibitem{Cox2004}
S.~J. Cox, ``Estimating the location and time course of synaptic input from
  multi-site potential recordings,'' \emph{J. of Computational Neuroscience},
  vol.~17, pp. 225--243, 2004.

\bibitem{Rinzel98}
J.~R. Rinzel and G.~B. Ermentrout, ``Analysis of neural excitability and
  oscillations,'' in \emph{Methods in {N}eural {M}odeling}, C.~Koch and
  I.~Segev, Eds.\hskip 1em plus 0.5em minus 0.4em\relax Cambridge, MA: MIT
  Press., 1998, pp. 135--169.

\bibitem{Tichavsky98}
P.~Tichavsk\'y, C.~H. Muravchik, and A.~Nehorai, ``{Posterior Cram\'er-Rao
  Bounds for Discrete-Time Nonlinear Filtering},'' vol.~46, no.~5, pp.
  1386--1396, May 1998.

\end{thebibliography}


\begin{IEEEbiography}[{\includegraphics[width=1in,height=1.25in,clip,keepaspectratio]{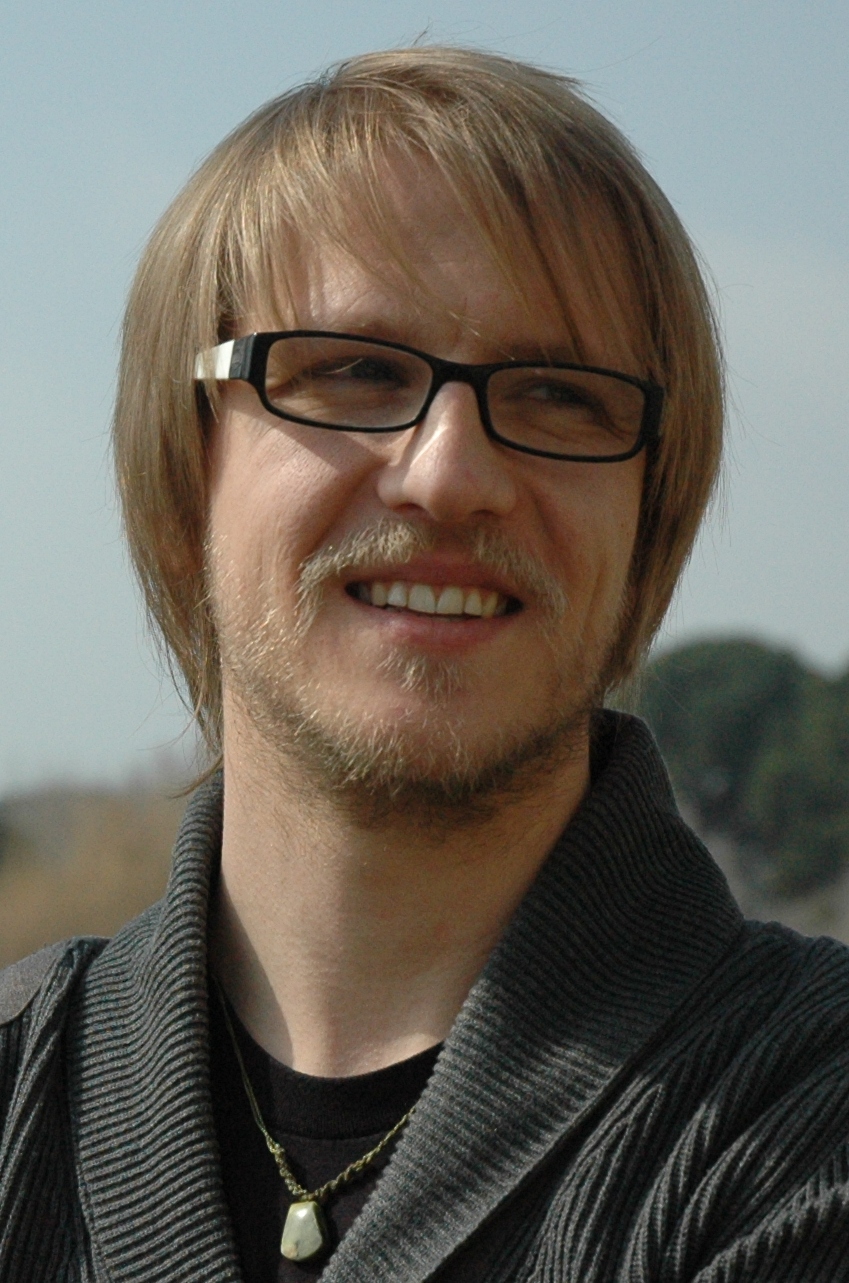}}]{Pau Closas} (S'04--M'10--SM'13) received the M.Sc. and Ph.D. in Electrical Engineering from the Universitat Polit\`ecnica de Catalunya (UPC) in 2003 and 2009, respectively. He also holds a M.Sc. degree in Advanced Mathematics and Mathematical Engineering from UPC since 2014. In 2003, he joined the Department of Signal Theory and Communications, UPC, as a Research Assistant. During 2008 he was Research Visitor at the Stony Brook University (SBU), New York, USA.

In September 2009 he joined the CTTC, where he currently holds a
position as a Senior Researcher and Head of the Statistical
Inference for Communications and Positioning Department. He has many
years of experience in projects funded by the European Commission,
Spanish and Catalan Governments, as well as the European Space
Agency in both technical and managerial duties. His primary areas of
interest include statistical and array signal processing, estimation
and detection theory, Bayesian filtering, robustness analysis, and
game theory, with applications to positioning systems, wireless
communications, and mathematical biology.

Pau Closas is Senior Member of IEEE, EURASIP, and ION. He was
involved in the organizing committees of EUSIPCO'11, IEEE IMWS'11,
IEEE RFID-TA'11, European Wireless'14, IEEE SSP'16, and IEEE
ICASSP'20 conferences. He is the recipient of the EURASIP Best PhD
Thesis Award 2014 and the $9^{th}$ Duran Farell Award for Technology
Research.
\end{IEEEbiography}

\begin{IEEEbiography}[{\includegraphics[width=1in,height=1.25in,clip,keepaspectratio]{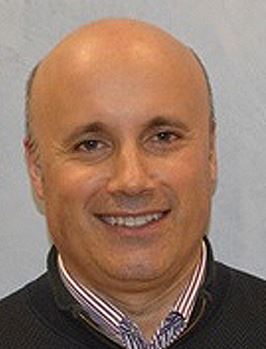}}]{Antoni Guillamon}
received the M.Sc. and Ph.D. in Mathematics (Applied) from the
Universitat Aut\`{o}noma de Barcelona (UAB) in 1989 and 1995,
respectively. He is Associate Professor of Applied Mathematics at
the Department of Mathematics, Universitat Polit\`{e}cnica de
Catalunya (UPC), since 1997. Currently, since 2011, he is also
serving as Deputy Director of the Centre de Recerca Matem\`{a}tica
and in the Monitoring Committee of the Barcelona Graduate School of
Mathematics.

He is an expert in dynamical systems theory (theory of limit cycles, study of invariant manifolds and bifurcations), both from analytical and computational perspectives. He has also contributed to the field of computational neuroscience, since the academic year 2000--01, when he was a Postdoctoral Researcher in the Courant Institute of Mathematical Sciences of the New York University (NYU), New York, USA. In this research field, he has been working in several problems: estimation of synaptic conductances, effects of synaptic depression in neuronal networks, bistable perception and synchronization problems in transient states. He was also co-ordinator of the first international conference on Mathematical Neuroscience (Neuromath'06), and has participated in several scientific committees of events related to this broad topic.
\end{IEEEbiography}

\end{document}